\documentclass[a4paper,12pt]{article}
\usepackage[cmex10]{amsmath}
\usepackage{latexsym,amssymb}
\usepackage[mathscr]{eucal}
\usepackage{amsxtra,amscd,amsthm,upref,layout,color}
\usepackage{calc}
\usepackage[final]{graphicx}
\def\refup#1{{$^{#1}$}}\def\refeq#1{(\ref{#1})}
\def\p{^{^{\prime}}}\def\pp{^{^{\prime\prime}}}

\def\min{{\rm {min}}}\def\max{{\rm{max}}}

\def\br{{\bf r}}

\def\oo{{\leavevmode\setbox0=\hbox{h}\dimen0=\ht0 \advance\dimen0
by-1ex\rlap{\raise0.47\dimen0\hbox{\char'27}}o}}
\def\bfPz{${\bf P_{0}  }$}\def\bfPu{${\bf P_{1}  }$}\def\bfPd{${\bf P_{2}  }$}
\def\begeq{\begin{equation}}
\def\endeq{\end{equation}}
\def\begdis{\begin{displaymath}}
\def\enddis{\end{displaymath}}

\def\cA{{\cal A}}\def\cB{{\cal B}}\def\cC{{\cal C}}\def\cD{{\cal D}}
 \def\cJ{{\cal J}} 
\def\cF{{\cal F}}  
\def\cL{{\cal L}}  
\def\cS{{\cal S}}\def\cT{{\cal T}}

\def\cDD{{\underline{\cD}(r)}}\def\cDDj{{\underline{\cD}_j(r)} }
\def\cDDo{{\underline{\cD}_o} }\def\Phio{{\underline{\Phi}_o} }
\def\gota{\mathfrak{a}}\def\gotA{\mathfrak{A}}
\def\gotb{\mathfrak{b}}\def\gotB{\mathfrak{B}}
\def\gotap{{\mathfrak{a}}\p}\def\gotAp{{\mathfrak{A}}\p}
\def\gotbp{{\mathfrak{b}}\p}\def\gotBp{{\mathfrak{B}}\p}
\def\gotC{{\mathfrak{C}}}
\def\gotD{{\mathfrak{D}}}
\def\gotF{{\mathfrak{F}}}
\def\gotP{{\mathfrak{P}}}
\def\gotf#1{{\mathfrak{f}}_{#1}}
\def\expcite#1{$^{[#1]}$}
\def\S1{{S$_1$}}

\def\arctan{\rm{arctan}}\def\arcos{\rm{arcos}}\def\arcsin{\rm{arcsin}}

\def\tht{\Theta_T(\cdots)}
\def\ellp{\ell\p}\def\cLp{\cL\p}
\def\oy{{\overline y}}\def\oY{{\overline Y}}\def\oX{{\overline X}}\def\oell{{\overline{\ell}}}\def\ocL{{\overline{\cL}}}
\def\oellp{{\overline{\ell\p}}}\def\ocLp{{\overline{\cL\p}}}
\def\ie{{\em i.e.}}\def\eg{{\em e.g.}}
\def\etal{{\em et al.}}
\def\phiC#1{$\underline {{\Phi_{#1}}^C}$}
\def\hw{{\hat \omega}} \def\hnu{{\hat {\nu}}}

\begin{document}
\title{The chord-length distribution  of a polyhedron}
\author{ 
{{ 
 Salvino Ciccariello$^{a,b,*}$
}}\\
%
  \begin{minipage}[t]{0.9\textwidth}
   \begin{flushleft}
\setlength{\baselineskip}{12pt}
{\slshape  {\footnotesize{
$^{a}$ Universit\`{a} di Padova, Dipartimento di Fisica {\em {G. Galilei}}, 
 Via Marzolo 8, I-35131 Padova, Italy, 
and \\
$^{b}$Universit\`{a}  {\em {C\`{a} Foscari}}, Department of Molecular Sciences and Nanosystems, 
Via Torino 155/B, I-30172 Venezia, Italy
}}}\\
 \footnotesize{$^*$salvino.ciccariello@unipd.it}
\end{flushleft}
\end{minipage}
}      
%
\date{\today  }   

\maketitle                        
\begin{abstract} \noindent 
The chord-length distribution function $[\gamma\pp(r)]$  of any 
bounded polyhedron has an elementary algebraic form, the expression of which 
changes in the different subdomains of the $r$-range. In each of these, 
 the $\gamma\pp(r)$ expression only involves, as transcendental contributions, 
inverse trigonometric functions of argument  equal to 
$R[r,\,\Delta_1]$, \,$\Delta_1$  being  the square root of a 
2nd-degree $r$-polynomial  and $R[x,y]$ a rational function.  
Besides, as  $r$ approaches 
one  boundary point ($\delta$) of each $r$-subdomain, the derivative of 
$\gamma\pp(r)$ can only show 
singularities of the forms $(r-\delta)^{-n}$ and $(r-\delta)^{-m+1/2}$ 
with $n$ and $m$ appropriate positive integers. Finally, the explicit algebraic 
expressions of the primitives are also reported.
\\    \\   


\end{abstract}
\vfill
\eject
{}{}
\subsection*{ 1. Introduction}  
Since long the correlation function (CF) has become a useful theoretical tool in 
almost  all scientific disciplines as, to mention just a few, the small-angle 
scattering\expcite{1,2}, the signal theory\expcite{3}, the pattern recognition 
theory\expcite{4}  and the stochastic geometry\expcite{5}. It also happens that its 
name changes with 
the discipline, since it is often  referred to as covariogram in the last two 
disciplines. Its use being so general, no wonder that it is studied both from 
the view point of practical applications and from  that of establishing 
the rigorous conditions that ensure its existence. In fact, given a  one 
dimensional physical quantity $\eta(x)$, its CF is defined as 
\begeq\label{1.1}
\gamma(r) =\lim_{L\to\infty}\frac{1}{2\,L}\int_{-L}^{L}\eta(x+r)\eta(x)dx.
\endeq
The definition clearly  involves a limit procedure with the consequent problem of 
establishing the conditions on $\eta(x)$ for the limit to exist. Wiener\expcite{6}  was 
one of the first mathematicians to afford this problem. In particular he got a 
result\expcite{6} that we like to mention for its generality and  elegance: "if $\gamma(r)$ 
exists for any $r$ and if it continuous at $r=0$ then it is everywhere continuous" (see, also, 
reference [7]). \\ 
In the case of small-angle scattering  theory it happens that $\eta(\br)$, 
the so-called scattering density fluctuation,  can fairly be looked at as a two value 
function and the  CF is defined as 
\begeq\label{1.2}
\gamma(r) =\lim_{V\to\infty}\frac{1}{4\,\pi\,V\langle \eta^2\rangle_V}
\int d\hw\int_{V}\eta(\br_1+r\hw)\eta(\br_1)dv_1.
\endeq
Here,  $dv_1$ is the volume infinitesimal element set at position $\br_1$,  
$\langle \eta^2\rangle_V$ denotes the mean value of function $\eta^2(\br)$ 
evaluated over the volume $V$,  and 
$\hw$ 
is a unit vector which can takes all possible orientations over which the first 
integral, accounting for  $(4\pi)^{-1}$ factor, evaluates the angular mean.  
Definition \refeq{1.2} ensures that $\gamma(0)=1$. To make clear the reason 
why definition \refeq{1.2} is related the covariogram of a geometrical body  
we first recall that the covariogram measures the angular average of the 
overlapping volume  of the body with its image, resulting by a translation 
of the body by $r\hw$. Assume that $\eta(\br)$ refers to a collection of 
particles, having the same shape and size and randomly distributed in 
the space,   and that the collection is dilute. Besides, let $\eta(\br)$ be 
equal to one inside the particles and to zero elsewhere.   The dilution 
and the randomness conditions allow us to approximate integral 
\refeq{1.2} by 
\begeq\label{1.3}
\gamma(r) \approx\lim_{V\to\infty}\frac{N}{4\,\pi\,V\langle \eta^2\rangle_V}
\int d\hw\int_{\underline{V_p}}\eta(\br_1+r\hw)\eta(\br_1)dv_1,
\endeq
where $N$ denotes the number of the particles inside the volume $V$ and 
$\underline{V_p}$ the spatial set (with volume $V_p$) occupied by a single 
particle. Since 
$\langle \eta^2\rangle_V$  is equal to $N\,V_p/V$,  
 equation \refeq{1.3},  in the limit $V\to\infty$, becomes 
\begeq\label{1.3}
\gamma_p(r) =\frac{1}{4\,\pi\,V_p}\int d\hw
\int_{\underline{V_p}}\eta(\br_1+r\hw)\eta(\br_1)dv_1,
\endeq
that is the  particle covariogram definition. The second derivative of this 
function, adapting to the present case the general expression obtained  
by Ciccariello \etal\expcite{8},  has the following integral expression
\begeq\label{1.4} 
\gamma\pp(r)=-\frac{1}{4\pi V}\int d\hw\int_{\cS}  dS_1 \int_{\cS} 
(\hnu_1\cdot\hw)(\hnu_2\cdot\hw)
\delta(\br_1+r\hw-\br_2)dS_2.
\endeq
Here, for simplicity, we omitted suffix $p$. Besides, $\cS$ denotes 
the boundary surface of the particle,  $\delta(\cdot)$ 
the three-dimensional Dirac function, $\hnu_1$  ($\hnu_2$) 
the unit normal to the infinitesimal surface element $dS_1$ ($dS_2$)
located at the point $\br_1$ ($\br_2$).  It is also assumed that 
$\cS$ is an orientable surface and that the considered unit normals 
point externally to the particle.  $\gamma\pp(r)$ also represents 
the probability density that, randomly tossing a stick of length $r$, 
the ends of the stick lie on $\cS$.   This property explains why 
$\gamma\pp(r)$ is investigated in the realm of stochastic geometry 
where $\gamma\pp(r)$ is often referred as chord length distribution 
(CLD). At the same time, the study of $\gamma\pp(r)$ is also 
relevant to the integral geometry\expcite{9}  that  
investigates the relations existing between the geometry of a 
body and some integrals over the latter.\\ 
These considerations  explain why efforts to get the CF or 
the CLD of particles with a well definite shape are valuable. So far 
we explicitly know the CFs of the sphere\expcite{1}, 
the cube\expcite{10}, the right parallelepiped\expcite{11}, 
the tetrahedron\expcite{12}, the octahedron\expcite{13} 
 and the circular cylinder\expcite{14,15}.  
For all these  shapes, except the cylindrical one,  
the CFs turn out to be simple algebraic functions, while the cylinder 
CF involves elliptical integral functions.  If we recall that the 
two-dimensional  CF of any  bounded plane polygon also has an 
algebraic form\expcite{16,17}  we are led 
to conjecture that the CF of any bounded polyhedron, whatever its 
shape, has an algebraic expression.\\
In this paper we show the truth of  a weaker form of this conjecture 
in so far it certainly applies to the CLD of any polyhedron. [As yet, 
we do not know if it also applies to the CF.] In fact, we  prove the 
following property:\\
{\bfPz} - {\em the CLD of any bounded polyhedron can be expressed 
in terms of elementary algebraic functions and inverse trigonometric 
functions depending  on rational functions of the two variables: $r$ 
and  $\sqrt{P_2(r)}$, with $P_2(r)$ equal to a 2nd degree 
$r$-polynomial}. \\ 
The plan of the paper is as follows. In the next section we show that 
a decomposition of the facets of the polyhedron into an union of triangles 
allows us to consider the integration superficial domains, present in 
\refeq{1.4}, as triangular ones. First we consider the case where the two 
triangles are not parallel. Section 3 shows that the six-dimensional 
integral \refeq{1.4}  can be converted in a two-dimensional one.  
Section 4 explicitly  performs  a further 
quadrature so as to convert  \refeq{1.4} into a one-dimensional integral. 
The values of this integral depends on the bounds of the integration 
domain that must be determined by reducing the inequalities present 
in the integrand definition. Section 5 performs the main steps of 
this task and section 6 
shows that the resulting integrand is a rational function $R(x,y)$ 
with $y$ equal to the square root of a second degree polynomial of 
the integration variable $x$. In this way property {\bfPz} is proved. 
Section 7 analyzes the case where the two integration triangles are 
parallel confirming that the aforesaid result holds also true in this case. 
Section 8 draws the final conclusions. The appendix reports the explicit 
algebraic expressions of the primitives which must be evaluated at the 
appropriate end-points of the last one-dimensional integral to get 
the explicit expression of the polyhedron CLD. 
\subsection*{2. Basic mathematical definitions}
Our task consists in evaluating integral \refeq{1.4} knowing that 
$\cS$ is the surface bounding a polyhedron of arbitrary shape  
having however a finite maximal chord. 
For any  polyhedron, the bounding surface $\cS$ is made of up of plane 
polygons $\cS_1,\,\cS_2,\,\ldots,,\cS_N$, so as to have 
$\cS=\cup_{i=1}^N \cS_i$. Consequently,  integral (\ref{1.4}) becomes   
\begeq\label{2.3} 
\gamma\pp(r)={\sum}_{\, i,j=1}^{,\,N}   g_{i,j}(r)
\endeq
with 
\begeq\label{2.4}
g_{i,j}(r)\equiv 
-\frac{1}{4\pi V}\int d\hw\int_{\cS_i}  dS_1 \int_{\cS_j} 
(\hnu_1\cdot\hw)(\hnu_2\cdot\hw)
\delta(\br_1+r\hw-\br_2)dS_2  
\endeq
\begin{figure}[h]
 {\includegraphics[width=10.truecm]{{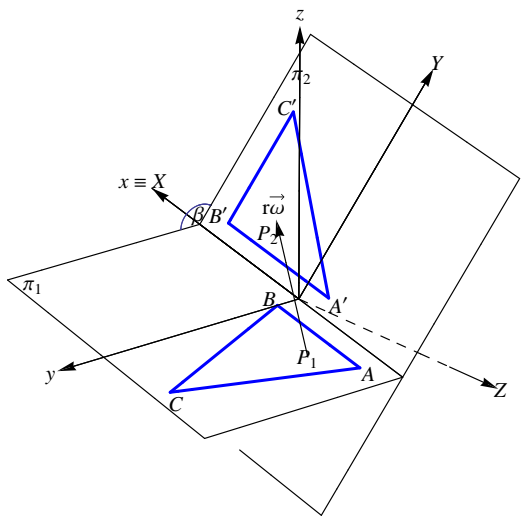}}}
\caption{\label{FigA1} { Typical configuration of two non-coplanar 
triangles and relevant Cartesian frame.       }}
\end{figure}
The prime on the summation symbol indicates that the cases  $i=j$ can 
there be omitted because $g_{i,i}(r)\equiv 0$. In fact, if $i=j$, the unit 
vectors  $\hnu_1$ and $\hnu_2$ are equal and the end points of $\br_1$ 
and $\br_2$ lie onto the considered facet's plane. 
Since the  Dirac function requires that $r\hw=\br_2-\br_1$, it follows 
that $\hw$ also lies  onto the aforesaid plane and is, therefore,  
orthogonal both to  $\nu_1$ and to $\nu_2$. Then the integrand of 
\refeq{2.4} vanishes and the property is proved. Thus, owing to equation 
(\ref{2.3}), the problem of evaluating $\gamma\pp(r)$ 
becomes  that of evaluating the $g_{i,j}(r)$s  with $i\ne j$. \\ 
Hereafter, for notational simplicity, we shall put $i=1$ and $j=2$. Let 
$\pi_1$ denote the plane on which $\cS_1$ lies and $\pi_2$ that 
relevant to $\cS_2$. Consider first the case where $\pi_1$ and $\pi_2$ 
intersect each other  along a line that we choose as the $x$ axis of a 
Cartesian orthogonal frame (see Fig. 1). Let $V_1,\,V_2,\,\cdots,\,V_M$ 
denote the vertices of the polygon $\cS_1$. We draw 
along each of the $V_l$s a straight line parallel to $x$. In this way, 
$\cS_1$ is divided into a set of trapezia, even though some of these can 
be simple triangles (see Fig. 2). Considering the only trapezia, each of 
these,  by considering one of its diagonals, splits into two triangles. By 
so doing, we have split $\cS_1$ into the union of $N_1$ triangles $\cT_i$ 
(each of these having one side parallel to axis $x$), 
\ie\  $\cS_1=\bigcup_{l=1}^{N_1}\cT_{l}$. A similar decomposition applies 
to $\cS_2$, \ie\ $\cS_2=\bigcup_{l=1}^{N_2}\cT\p_{l}$. 
Once we use the above two decompositions of $\cS_1$ and $\cS_2$ into 
equation (\ref{2.4}),  $g_{1,2}(r)$ becomes a  sum of terms that have the 
same structure of (\ref{2.4}) with the only change that $\cS_1$ and 
$\cS_2$ are now simple triangles. Hence, our task is that of evaluating 
the following integral expression 
\begeq\label{2.5}
g(r)\equiv 
-\frac{1}{4\pi V}\int d\hw\int_{\cT}  dS_1 \int_{\cT\p} 
(\hnu_1\cdot\hw)(\hnu_2\cdot\hw)
\delta(\br_1+r\hw-\br_2)dS_2,
\endeq
where, for greater notational simplicity, we omitted indices on the 
considered triangles and on  the $g(r)$ symbol. \\ 
\subsection*{3.  Reduction of integral (\ref{2.5}) to a two-dimensional integral} 
To evaluate integral (\ref{2.5}), we first define the most convenient 
Cartesian frame.  To thi aim we refer to Fig. 1 that shows the two triangles 
$\cT=ABC$ and   $\cT\p=A\p B\p C\p$.  As already anticipated, we choose the 
$x$-axis along the intersection line of the planes containing the two triangles. 
At first, we arbitrarily choose  one of the two possible orientations for $x$. 
Then, the oriented $y$-axis is chosen perpendicularly to $x$ and in such a way 
that $\cT$ fully lies in the region $y\ge 0$. The oriented $z$-axis 
is chosen perpendicularly to $x$ and $y$  and in such a 
way that the resulting system $Oxyz$ be right-handed. Now, we anti-clock-wisely 
rotate around $x$ the half-plane containing $\cT$ by an angle $\beta$ till 
it superposes to the half-plane containing $\cT\p$. We always can choose 
$Oxyz$ in such a way that $0<\beta<\pi$. In fact, if proceeding as just said, 
we find that $\beta$ exceeds $\pi$, then we choose for $x$ and $z$ the 
opposite directions and in this way the resulting $Oxyz$ is still right-handed 
and the resulting $\beta$ obeys to $0<\beta<\pi$. 
We also observe that it is not restrictive to assume that both $\hnu_1$ and 
$\hnu_2$ point towards the interior of the dihedral angle $\beta$ so as to 
have
\begeq\label{3.1}
\hnu_1=(0,\,0,\,1)\quad{\rm and}\quad  \hnu_2=(0,\,\sin\beta,\,-\cos\beta).
\endeq
In fact, whenever one or both of these conditions were not realized, we  
change the direction(s) of the normal(s) that points (point) outside and, 
at the end of the integral evaluation, we change the sign of the result 
in the only case where one normal has undergone a change of direction.\\ 
With respect to the  chosen $Oxyz$ frame, the components of 
$\br_1$ are $(x_1,\,y_1,\,0)$ that we find it more convenient to rename 
as $(x,\,y,\,0)$. It is also convenient to introduce a further  
cartesian frame $OXYZ$ having axis $X$ coinciding with $x$, axis $Y$ 
orthogonal to $X$ and oriented in such  a way that  triangle $\cT\p$ 
lies in the region $Y\ge 0$, and, finally, $Z$ orthogonal to both $X$ 
and $Y$ (see Fig. 1) and with  orientation such as to ensure the 
right-handedness of $OXYZ$. With respect to this frame the components of 
$\br_2$ are $(X,\,Y,\,0)$. These, converted to the $Oxyz$ frame, become 
$(X,\,Y\cos\beta,\,Y\sin\beta)$.   Finally, choosing $z$ as polar 
axis and axis $x$ as origin of the longitudinal angle,  the $Oxyz$ 
components  of $\hw$ are 
$\hw=(\sin\theta\cos\varphi,\,\sin\theta\sin\varphi,\,\cos\theta)$. 
Besides, we have $d\hw=\sin\theta d\theta d\varphi $, $dS_1=dxdy$ and 
$dS_2=dXdY$. By these definitions  follows that  the Dirac function 
requires  the fulfillement of the following three equalities:  
\begin{eqnarray}\label{3.2}
X&=&x+ r \sin\theta\cos\varphi,\\
Y\cos\beta&=& y+r\sin\theta\sin\varphi,\label{3.3}\\
Y\sin\beta&=& r\cos\theta.\label{3.4}
\end{eqnarray}
We solve these equations with respect to $Y,\, X$ and $y$ 
and denote the solutions as 
\begin{eqnarray}
&& {\oY}\equiv {\oY}(r,\theta)\equiv r\cos\theta/\sin\beta,\label{3.6}\\
&& {\oy}\equiv {\oy}(r,\theta,\varphi)\equiv 
r\, \cot\beta\, \cos\theta-r\,\sin\theta\,\sin\varphi,\label{3.7}\\
&&{\oX}\equiv {\oX}(x,r,\theta\varphi)\equiv x+ 
r \sin\theta\cos\varphi.\label{3.8}
\end{eqnarray}
The Dirac function allows us to explicitly perform the integrations 
with respect to $X,\, Y$ and $y$  in equation (\ref{2.5}). We obtain 
\begeq\label{3.9}
g(r)=-\frac{1}{4\pi V\sin\beta}\int d\hw \int dx (\hnu_1\cdot\hw)
(\hnu_2\cdot\hw)\tht.
\endeq
Here $\tht$ denotes a product of Heaviside functions that, as it will 
be expounded below,  ensures that solutions  (\ref{3.6})-(\ref{3.7})  
fall inside triangles $\cT$ and $\cT\p$. 
In fact, the condition that $\br_2\in\cT\p$ implies that:
\begeq\label{3.10}
Y_m\le Y\le Y_M,\quad{\rm and}\quad \ellp(Y)\le X \le \cLp(Y),
\endeq
where $ Y_m$ and $Y_M$ respectively denote the $Y$ ordinate  of 
the side $A\p B\p$ points and that of vertex $C\p$ (see  Fig.1), while 
\begeq\label{3.11}
\ellp(Y)=X_{A\p}+\frac{X_{C\p}-X_{A\p}}{Y_M-Y_m}(Y-Y_m)\equiv 
\gotap+\gotbp Y 
\endeq 
is the $OXY$ equation of  the side  $A\p C\p$   and 
\begeq\label{3.12}
\cLp(Y)=X_{B\p}+\frac{X_{C\p}-X_{B\p}}{Y_M-Y_m}(Y-Y_m)\equiv 
\gotAp+\gotBp Y 
\endeq
that of the side $B\p C\p$.  The rightmost sides of the above two 
equations represent  a more  compact notation of the equations 
and the values of constants $\mathfrak{a\p,\,b\p,\,A\p}$ 
 {\rm and} $\gotBp$ are easily obtained by the polynomial identity 
principle. We note that, whenever $\cT\p$ have a form such that 
$C\p$ were closer to axis $x$  than the side $A\p B\p$, then $Y_m$ 
and $Y_M$ are the $Y$-ordinates of $C\p$ and $A\p B\p$, 
respectively, and $Y_m$ and $Y_M$ must be interchanged in the 
middle sides of \refeq{3.11}  and \refeq{3.12}. Thus, in general, 
$Y_m$ ($Y_M$) denotes the smallest 
(largest) $Y$-ordinate of the points of $\cT\p$.  
Quite similarly, the $(x,y)$ ordinates of the points of $\cT$ 
obey the inequalities 
\begeq\label{3.13}
y_m\le y\le y_M,\quad{\rm and}\quad \ell(y)\le x \le \cL(y),
\endeq
where, now, $\ell(y)\equiv\gota+\gotb y$ and 
$\cL(y)\equiv\gotA+\gotB\, y$ respectively are the $Oxy$ equations of 
sides $AC$ and $BC$ of triangle $\cT$ and  are defined 
quite similarly to equations (\ref{3.11}) and (\ref{3.12}).  
It follows that  the solutions of equations (\ref{3.2})-(\ref{3.4}) 
must obey  inequalities  (\ref{3.10}) and the first of (\ref{3.13}), 
\ie : 
\begin{eqnarray}\label{3.14}
&& Y_m\le {\oY}(r,\theta) \le Y_M, \\
&&y_m\le {\oy}(r,\theta,\varphi)\le y_M,\label{3.15}\\
&&\ellp({\oY})\le {\oX}(x,r,\theta,\varphi)\le 
\cLp({\oY}).\label{3.16}
\end{eqnarray}     
Owing to equation (\ref{3.8}), inequality  (\ref{3.16}) can be 
converted into an inequality for $x$ since it can be written as 
\begeq\label{3.17}
\oellp(r,\theta,\varphi)\le x \le \ocLp(r, \theta,\varphi),
\endeq
where we have put 
\begeq\label{3.18}
\oellp=\oellp(r,\theta,\varphi)\equiv \ellp(\oY)-r\sin\theta\,\cos\varphi= 
\gotap+r(\gotbp\frac{\cos\theta}{\sin\beta}-\sin\theta\,\cos\varphi),
\endeq
and
\begeq\label{3.19}
\ocLp=\ocLp(r,\theta,\varphi)\equiv \cLp(\oY)-r\sin\theta\,\cos\varphi=
\gotAp+r(\gotBp\frac{\cos\theta}{\sin\beta}-\sin\theta\,\cos\varphi).
\endeq 
Combining inequalities (\ref{3.17}) and the second of  (\ref{3.13}), 
evaluated at $y=\oy$, we obtain  the final inequalities for variable 
$x$, namely 
\begeq\label{3.20}
{\rm max}[\ell(\oy), \oellp(r,\theta,\varphi)]\le x 
\le {\rm min}[\cL(\oy), \ocLp(r,\theta,\varphi)],
\endeq 
where we have put 
\begeq\label{3.21}
\oell=\ell(\oy)=\mathfrak{a}+\mathfrak{b}\, r\, 
(\cot\beta\cos\theta-\sin\theta\sin\varphi)
\endeq
and 
\begeq\label{3.22}
\ocL=\cL(\oy)=\mathfrak{A}+\mathfrak{B}\, r\, 
(\cot\beta\cos\theta-\sin\theta\sin\varphi).
\endeq 
We explicit note a property that will often be used later 
under the name of substitution property: 
the substitution $\{\gota\to \gotA,\, \gotb\to \gotB\}$ 
and its inverse 
$\{\gotA\to \gota,\,\gotB\to \gotb\}$ respectively 
transform $\oell$  into $\ocL$ and $\ocL$ into $\oell$. 
Similarly, $\{\gotap\to \gotAp,\,\gotbp\to \gotBp \}$ 
transforms $\oellp$ into $\ocLp$ and 
$\{\gotAp\to \gotap,\,\gotBp\to \gotbp \}$  transforms 
$\ocLp$ into $\oellp$. \\
Using relation \refeq{3.20}, integral (\ref{3.9}) can be further 
integrated with respect to $x$ to yield 
\begeq\label{3.23}
g(r)=-\frac{1}{4\pi V\sin\beta}\int d\hw  (\hnu_1\cdot\hw)
(\hnu_2\cdot\hw)\bigl({\rm min}[\ocL, \ocLp]-
{\rm max}[\oell, \oellp]\bigr)\tht,
\endeq
where $\tht$ denotes that the integration variables $\theta$ and 
$\varphi$ must  be restricted  to  the subdomain $\cDD$ of $\cD_o$ 
where inequalities \refeq{3.14}, (\ref{3.15}) and (\ref{3.20}) are obeyed. 
[Domain  $\cDDo$ denotes  the outset integration domain defined as  
$\cDDo\equiv \{\theta,\,\varphi\,| 0<\theta<\pi;\,\, \varphi\in\Phio\}
 $,  with $\Phio\equiv\{\varphi\,|  -\pi/2<\varphi<3\pi/2\}$. 
The last  $\varphi$ choice will turn out to be convenient later.] \\ 
The fact that the CLD of a polyhedron is a sum of contributions having 
the form of equation \refeq{3.23} allows us to state the property:\\
{\bfPu} - {\em the CLD of any polyhedron can alway be written as a 
sum of two-dimensional integrals}. 
\subsection*{4. Reduction of the CLD to a single quadrature}
The shape of $\cDD$ can exactly be determined by reducing 
inequalities \refeq{3.14}, \refeq{3.15} and \refeq{3.20}, a task
 in principle simple but in practice long and boring to 
be  carried out.  We shall later do some steps in this 
reduction analysis. For the moment, we only note that $\cDD$ 
changes its shape with $r$ and that it may consist of disjoined 
sets  (see, \eg, Ref.\,[11]). 
In any case, $\cDD$ can be partitioned 
into the union of smaller sets  $\cDDj$ with $j=1,\ldots,M_{\cD}$ 
such that each $\cDDj$ can be written as 
$\{\theta,\,\varphi|\,\theta_{j,m}<\theta<\theta_{j,M},\,
\phi_{j,m}(\theta)<\varphi<\phi_{j,M}(\theta)\}$, 
where the dependence of the bounds 
$\theta_{j,m},\ \theta_{j,M},\ \phi_{j,m}(\theta)$ and 
$\phi_{j,M}(\theta)$  on $r$ is omitted 
for simplicity. The determination of these bounds 
rests upon the reduction of  the aforesaid three inequalities.   
Hence, in a given $\cDDj$, we know if 
${\rm min}[\ocL, \ocLp]$ is equal to $\ocL$ or to $\ocLp$ and if  
${\rm max}[\oell, \oellp]$ is equal to $\oell$ or to $\oellp$ and 
that inequality  ${\rm min}[\ocL, \ocLp]>{\rm max}[\oell, \oellp]$ 
is  obeyed. \\ 
From these considerations we draw the following general property:\\ 
{\bfPd } - {\em the determination of the CLD of a polyhedron always 
reduces to a quadrature problem.}\\   
To prove this statement, we recall that the determination of the 
CLD reduces to a sum of contributions of the form \refeq{2.4} where 
the integration domains are simple triangles and that all  these 
contributions  have the form of the two-dimensional integral \refeq{3.23}. 
It is now convenient to introduce the following quantities   
\begeq\label{4.1}
\Psi(\theta,\varphi)\equiv  
(\hnu_1\cdot\hw)(\hnu_2\cdot\hw) = 
\cos\theta\,[\sin\beta\,\sin\theta\,\sin\varphi-\cos\beta\,\cos\theta],
\endeq 
\begin{eqnarray}
& & \cF_1(r,\theta,\varphi)\equiv 
\Psi(\theta,\varphi)\oell(r,\theta,\varphi) =\label{4.2}\\
&&\quad\quad -\cos\beta\,\cos^2\theta(\gota+\gotb\,r\, 
\cot\beta\cos\theta)+\sin\varphi\,\sin\theta\,
\cos\theta\times \nonumber \\
&&\quad\quad  (\gota\,\sin\beta+2\,\gotb\,r\,
\cos\beta\,\cos\theta)-\gotb\,r\,\sin\beta\,\sin^2\varphi\,
\cos\theta\,\sin^2\theta,\nonumber
\end{eqnarray}
\begin{eqnarray}
&&\cF_2(r,\theta,\varphi)\equiv \Psi(\theta,\varphi)
\ocL(r,\theta,\varphi)=
\cF_1(r,\theta,\varphi)\big|_{(\gota\to\gotA,\, \gotb\to\gotB)},
\quad\quad\quad\label{4.3}
\end{eqnarray}
\begin{eqnarray}
& &\cF_3(r,\theta,\varphi)\equiv 
\Psi(\theta,\varphi)\oellp(r,\theta,\varphi)=\label{4.4}\\
& &\quad\quad  - (\gotap\, \cos\beta+\gotbp\,r\,\cot\beta\,\cos\theta)
\cos^2\theta + 
r\,\cos\beta\,\cos\varphi\,\sin\theta\,\cos^2\theta+\nonumber\\
& &\quad\ \  \sin\varphi\,\cos\theta\,\sin\theta(\gotap\,\sin\beta + 
\gotbp\,r\,\cos\theta)
-r\,\sin\beta\,\cos\varphi\,\sin\varphi\,\cos\theta\,\sin^2\theta,\nonumber
\end{eqnarray}
\begin{eqnarray}
& &\cF_4(r,\theta,\varphi)\equiv \Psi(\theta,\varphi)\ocLp(r,\theta,\varphi)=
\cF_3(r,\theta,\varphi)\big|_{(\gotap\to\gotAp,\, \gotbp\to\gotBp)}.
\quad\quad\quad\quad\quad\quad\label{4.5}
\end{eqnarray}
Here, equation \refeq{4.1} follows from equation (\ref{3.1}) and the $\hw$ 
expression reported above equation \refeq{3.2} while the 
right hand sides of \refeq{4.2}-\refeq{4.5} follow  from equation (\ref{4.1}) and  
definitions \refeq {3.21}, \refeq {3.22},  \refeq{3.18} and \refeq{3.19}. 
Besides, on  the rightmost sides of \refeq{4.3} and \refeq{4.5} we used the 
substitution property reported below equation \refeq{3.22}. 
Recalling how the $\cDDj$s have been defined and using definitions 
\refeq{4.2}-\refeq{4.5},   after putting  
\begeq\label{4.7}
\cJ_{k}(r,\theta,\varphi)\equiv \int \cF_k(r,\theta,\varphi)d\varphi,
\quad k=1,\ldots,4,
\endeq 
equation (\ref{3.23}) becomes  
\begin{eqnarray}\label{4.6}
&&g(r)=-\sum_{j=1}^{M_{\cD}}\frac{1}{4\pi V\sin\beta}
\int_{\theta_{j,m}}^{\theta_{j,m}}
\sin\theta d\theta\times \\
 &&\quad\quad\quad\quad  \Big[\Big(\cJ_{a_{j}}(r,\theta,\phi_{j,M}(\theta))-
\cJ_{b_{j}}(r,\theta,\phi_{j,M}(\theta))\Big)-\nonumber \\
&& \quad\quad\quad\quad\quad\quad   \Big(\cJ_{a_{j}}(r,\theta,\phi_{j,m}(\theta))-
\cJ_{b_{j}}(r,\theta,\phi_{j,m}(\theta))\Big)\Big],\nonumber
\end{eqnarray}
where the values of indices $a_j$ and  $b_j$ are determined by the set 
$\cDDj$. Note that   $a_j$ is equal to 2 or 4 and index $b_j$ to 1 or 3.  
Definitions  \refeq{4.2}-\refeq{4.5} make it evident that 
the four $\cJ_{k}(r,\theta,\varphi)$ functions are explicitly known 
because the integrands  are polynomial  
functions  in the variables   $\sin\varphi$ and 
$\cos\varphi$..  Then equation \refeq{4.6} proves  that 
property  {\bfPd} is true. Moreover, relation \refeq{4.6} shows that if 
\begeq\label{4.8}
\int   \cJ_{k}(r,\theta,\phi_B(\theta))\,\sin\theta\,d\theta 
\endeq 
is an algebraic function for any $k=1,\ldots,4$ and 
for $\phi_B(\theta)$ equal to one  the possible  $\phi_{j,M}(\theta)$s or 
$\phi_{j,m}(\theta)$s, then $g(r)$ also is an algebraic function 
and property {\bfPz} is proved. 
\subsection*{5. Reduction of  inequalities \refeq{3.14}, \refeq{3.15} 
and \refeq{3.20} }
Hence, the proof of {\bfPz} amounts to prove that integral \refeq{4.8}  is an 
algebraic function.  
To this aim we need first to know how $\phi_{j,m}$ and $\phi_{j,M}$ depend 
on $\theta$ and  we must therefore elaborate inequalities (\ref{3.14}), 
\refeq{3.15} and (\ref{3.20}). \\
Inequality (\ref{3.14}) only concerns variable $\theta$  and yields 
\begeq\label{5.1} 
Y_m\sin\beta/r<\cos\theta < Y_M\sin\beta/r.
\endeq
This inequality only exists if $r>Y_m\sin\beta$. Thus, we see that the 
inequality reduction  also restricts the range of the acceptable 
$r$ values. Assuming the aforesaid constraint on $r$ obeyed, inequality 
\refeq{5.1} restricts variable $\theta$ to lie within the interval 
${\underline \Theta_I}$ defined as 
\begeq\label{5.2}
{\underline \Theta_I}\equiv[\arcos( Y_M\sin\beta/r),\,\arcos( Y_m\sin\beta/r)],
\endeq 
that must be interpreted  {\em cum grano salis} in the sense that the 
left bound must be set equal to zero whenever $Y_M\sin\beta/r>1$. 
We observe that ${\underline \Theta_I}$ is a subset of 
the interval $[0,\,\pi/2]$ so that quantities $\cos\theta$ and 
$\sin\theta$, encountered 
in the following analysis, always are non-negative.\\
Inequality \refeq{3.15}, solved with respect to $\varphi$, yields 
\begeq\label{5.3} 
\frac{\cA_{II,L}+\cB_{II}\cos\theta}{\sin\theta}<\sin\varphi<
\frac{\cA_{R,II}+\cB_{II}\cos\theta}{\sin\theta}
\endeq
with
\begeq\label{5.3a}
 \cA_{II,L}\equiv -y_M/r, \quad \cA_{R,II}\equiv -y_m/r\quad{\rm and} 
\quad\cB_{II}\equiv \cot\beta.
\endeq 
(Hereafter, subscripts L and R will always refer to left and right bounds.) 
For inequality \refeq{5.3} to exist it is necessary that 
\begeq\label{5.3b}
 ({\cA_{II,L}+\cB_{II}\cos\theta})/{\sin\theta}\le 1\quad{\rm and}\quad
 ({\cA_{R,II}+\cB_{II}\cos\theta})/{\sin\theta}\ge - 1.
\endeq
The  last two inequalities only constraint variables $\theta$ and $r$ and  
we shall not further analyze their implications because  we are 
only interested in the bounds on variable $\varphi$. Put 
\begin{eqnarray}\label{5.3c}
\phi_{II,L,1}&\equiv& \arcsin
\Big( \frac{\cA_{II,L}+\cB_{II}\cos\theta}{\sin\theta}\Big)
\end{eqnarray}
and 
\begin{eqnarray}\label{5.3d}
\phi_{II,R,1}&\equiv& \arcsin
\Big( \frac{\cA_{R,II}+\cB_{II}\cos\theta}{\sin\theta}\Big),
\end{eqnarray}
then inequality \refeq{5.3}  is equivalent to the validity of the following two 
inequalities
\begin{eqnarray}\label{5.3e}
 \phi_{II,L,1} \le &\varphi&\le \phi_{II,R,1},\\
\phi_{II,L,2}\equiv\pi-\phi_{II,R,1} \le & \varphi&\le
 \phi_{II,R,2}\equiv\pi-\phi_{II,L,1}.     \label{5.3f}
\end{eqnarray}
[Similarly to the remark reported below equation \refeq{5.2}, 
bounds $\phi_{II,L,1}$ and $\phi_{II,R,1}$ 
must respectively be set equal to $-\pi/2$ if 
$({\cA_{II,L}+\cB_{0}\cos\theta})/{\sin\theta}<-1$ and to 
$\pi/2$ if 
$({\cA_{II,R}+\cB_{II}\cos\theta})/{\sin\theta}>1$. Similar 
substitutions will be understood in all the following 
relations involving the $\arcsin$ function whenever the 
latter argument is smaller than -1 or greater than 1.]  
Hence, inequality \refeq{3.15} requires that  variable $\varphi$ 
be always confined to the set $\underline{\Phi_{II}}$ defined as 
\begeq\label{5.3g} 
\underline{\Phi_{II}}\equiv [\phi_{II,L,1},\, \phi_{II,R,1}]\cup 
[\phi_{II,L,2},\, \phi_{II,R,2}].
\endeq
We turn now to the analysis of the third inequality, \ie\ condition 
\refeq{3.20}.    This  involves the following cases:
\begin{eqnarray}
\oell& < &\oellp\, <\, \ocL\,<\, \ocLp,\label{5.3A}  \\
\oell& < &\oellp\, <\, \ocLp\,<\, \ocL,\label{5.3B}  \\
\oellp& < &\oell\, <\, \ocL\,<\, \ocLp,\label{5.3C}  \\
\oellp& < &\oell\, <\, \ocLp\,<\, \ocL.\label{5.3D}  
 \end{eqnarray}
These inequalities are reduced with respect to $\varphi$
also requiring   that  $\varphi\in\underline{\Phi_{II}}$ for 
inequality \refeq{3.15} to be obeyed.  \\  
We begin by considering the leftmost inequality of  \refeq{5.3A} 
and \refeq{5.3B}.  We first determine the 
$\varphi$ range, momentarily denoted by $\underline{\Phi}$, 
where it results $\ell(\oy) < \oellp(r,\theta,\varphi)$. 
Owing to definitions \refeq{4.2} and \refeq{4.3}, which associate $\oell$ 
to $\cF_1$ and  $\oellp$ to $\cF_3$, this case will be referred 
to as the "1\,3" case. Using definitions \refeq{3.21}  and 
\refeq{3.18}, inequality $\ell(\oy) <\oellp(r,\theta,\varphi)$ 
can be written as 
\begeq\label{5.4}
\gota-\gotap +r\cos\theta(\gotb\,\cot\beta-
\frac{\gotbp}{\sin\beta})<
r\sin\theta(\gotb\sin\varphi-\cos\varphi).
\endeq
Putting 
\begeq\label{5.5}
\cos\gotf{13} \equiv \gotb/\sqrt{1+\gotb^2}, \quad\quad 
\sin\gotf{13} \equiv 1/\sqrt{1+\gotb^2},
\endeq  
and 
\begeq\label{5.8}
\cA_{13}\equiv \frac{\gota-\gotap}{r\sqrt{1+\gotb^2}},
\quad\quad \cB_{13}\equiv 
\frac{\gotb\,\cos\beta-\gotbp}{\sin\beta\,\sqrt{1+\gotb^2}}, 
\endeq
inequality \refeq{5.4} takes a form similar to \refeq{5.3}, \ie 
\begin{eqnarray}\label{5.6}
\sin(\varphi-\gotf{13}) &>&
\frac{\cA_{13} +\cB_{13}\cos\theta}{\sin\theta}. \label{5.6}
\end{eqnarray} 
We conclude that $\oell<\oellp$ is obeyed within the interval 
\begeq\label{5.9}
\underline{\Phi_{13}}=[\phi_{13,L},\, \phi_{13,R}],
\endeq 
with
\begin{eqnarray}\label{5.10a}
\phi_{13,L}&\equiv&\gotf{13}+\arcsin\Big(
\frac{\cA_{13} +\cB_{13}\cos\theta}{\sin\theta}\Big),\\ 
\phi_{13,R}&\equiv& \pi +
\gotf{13}-\arcsin\Big(
\frac{\cA_{13} +\cB_{13}\cos\theta}{\sin\theta}\Big).\label{5.10b}
\end{eqnarray}
From this result follows that the leftmost inequality of \refeq{5.3C} 
and \refeq{5.3D} , \ie\ $\oell>\oellp$, is obeyed 
within the set  \phiC{13} that is the 
complement of $\underline{\Phi_{13}}$ to the set  
$\Phio$. In this way, the analysis of case "13" is accomplished.\\
We analyze now the inequality related to the rightmost sides of 
\refeq{5.3A} and \refeq{5.3C}, \ie\ the condition $\ocL\,<\,\ocLp$. 
This  will be referred to as the "24" case owing to definitions 
\refeq{4.3} and \refeq{4.5}.  The  rightmost sides of these  definitions  
show that inequality $\ocL\,<\,\ocLp$ follows from $\oell<\oellp$ 
once we apply to this the transformation $(\gota\to\gotA,\, 
\gotb\to\gotB,\,\gotap\to\gotAp,\, \gotbp\to\gotBp)$. 
Then, we conclude that inequality $\ocL\,<\,\ocLp$ is obeyed within 
the $\varphi$ interval $\underline{\Phi_{24}}$ that 
is obtained  applying the previous transformation  to 
$\underline{\Phi_{13}}$, \ie
\begeq\label{5.11}
\underline{\Phi_{24}} = 
\underline{\Phi_{13}}_{(\gota\to\gotA,\, 
\gotb\to\gotB,\,\gotap\to\gotAp,\, \gotbp\to\gotBp)}.
\endeq
The inequality $\ocL\,>\,\ocLp$, related to the rightmost sides of 
\refeq{5.3B} and \refeq{5.3D}, is fulfilled within the set  \phiC{24}   
that is   the complementary set of $\underline{\Phi_{24}}$ to 
$\Phio$ as well as the set obtained by applying the substitution 
$(\gota\to\gotA,\, \gotb\to\gotB,\,\gotap\to\gotAp,\, 
\gotbp\to\gotBp)$  to \phiC{13}. In this way, 
the analysis of case "24" also is fully accomplished.\\ 
To complete the reduction of inequalities \refeq{5.3A}-\refeq{5.3D} 
we must analyze the inequalities present in the middle of each of 
them.\\     
 We start by considering \refeq{5.3A} where the middle 
inequality is  $\oellp\,<\,\ocL$ that, owing to definitions \refeq{3.21}  
and \refeq{3.19},  will be referred to as case "23". It is more convenient 
to analyze first the opposite inequality, \ie\ $\oellp\,>\,\ocL$.  
This reads 
\begeq\label{5.13}
r\,\sin\theta(\gotB\,\sin\varphi-\cos\varphi) > 
\gotA-\gotap+r\,\cos\theta(\gotB\,\cot\beta-\gotbp\,\csc\beta).
\endeq
Putting 
\begeq\label{5.14}
\cos\gotf{23} \equiv \gotB/\sqrt{1+\gotB^2},\quad\quad 
\sin\gotf{23} \equiv 1/\sqrt{1+\gotB^2},
\endeq 
\begeq\label{5.16}
\cA_{23}\equiv \frac{\gotA-\gotap}{r\,\sqrt{1+\gotB^2}}\quad{\rm and}\quad 
\cB_{23}\equiv\frac {\gotB\,\cos\beta-\gotbp}{\sin\beta\,\sqrt{1+\gotB^2}},
\endeq
inequality \refeq{5.13} can be written in a form similar to 
\refeq{5.6}, \ie\   as 
\begeq\label{5.15}
\sin(\varphi-\gotf{23}) >
\frac{\cA_{23} +\cB_{23}\cos\theta}{\sin\theta}. 
\endeq
Thus, inequality $\oellp>\ocL $ is obeyed within the 
interval 
\begeq\label{5.17}
\underline{\Phi_{23}}=[\phi_{23,L},\, \phi_{23,R}]
\endeq 
with 
\begin{eqnarray}\label{5.18}
\phi_{23,L}&\equiv&\gotf{23}+\arcsin\Big(
\frac{\cA_{23} +\cB_{23}\cos\theta}{\sin\theta}\Big),\\ 
\phi_{23,R}&\equiv& \pi +
\gotf{23}-\arcsin\Big(
\frac{\cA_{23} +\cB_{23}\cos\theta}{\sin\theta}\Big),\label{5.19}
\end{eqnarray}
while the inequality $\oellp<\ocL $ is obeyed within 
$\underline{{\Phi^C}_{23}}$, the complementary set of 
$\underline{\Phi_{23}}$  to $\Phio$. We conclude that 
relation \refeq{5.3A} is obeyed  within the 
$\varphi$-set:  
\begeq\label{5.19a}
\underline{\Phi_{A}}\equiv\underline{\Phi_{13}}\cap\underline{{\Phi^C}_{23}}\cap
\underline{\Phi_{24}}\cap\underline{\Phi_{II}}, 
\endeq   
We pass now to  analyze inequality $\oellp\,<\,\ocLp$ present in the middle of 
relation \refeq{5.3B}.  This inequality will be referred to as the "3\,4" case. 
Since it  reads 
\begeq\label{5.19b}
r\,(\gotBp-\gotbp)\,\cos\theta >(\gotap-\gotAp)\sin\beta,
\endeq 
it puts no constraints on $\varphi$. Thus relation \refeq{5.3B} is obeyed 
within the $\varphi$-set
\begeq\label{5.19c}
\underline{\Phi_{B}}\equiv\underline{\Phi_{13}}\cap
\underline{{\Phi^C}_{24}}\cap\underline{\Phi_{II}}.
\endeq  
We  reduce now  inequality $\oell\,<\,\ocL$, present in the middle of 
relation \refeq{5.3C}. This case is referred to as the "1\,2" case. The inequality reads 
\begeq\label{5.19d}
r\,(\gotb-\gotB)\,\sin\theta\,\sin\varphi >
(\gota-\gotA)+r\,(\gotb-\gotB)\cot\beta\,\cos\theta. 
\endeq 
Assuming that $\gotb>\gotB$, and putting 
\begeq\label{5.19e}
\cA_{12}\equiv \frac{\gota-\gotA}{r\,(\gotb-\gotB)},\quad 
\cB_{12}\equiv \cot\beta,\quad
\endeq
the set of the allowed $\varphi$ values is 
\begeq\label{5.24}
\underline{\Phi_{12}}=[\phi_{12,L},\, \phi_{12,R}]
\endeq 
with 
\begin{eqnarray}\label{5.25}
\phi_{12,L}&\equiv&\arcsin\Big(
\frac{\cA_{12} +\cB_{12}\cos\theta}{\sin\theta}\Big),\\ 
\phi_{12,R}&\equiv& \pi -\arcsin\Big(
\frac{\cA_{12} +\cB_{12}\cos\theta}{\sin\theta}\Big),\label{5.26}
\end{eqnarray}
Whenever $\gotb<\gotB$, inequality \refeq{5.19d} becomes 
\begeq\label{5.27}
\sin\varphi <
\frac{(\gota-\gotA)/[r\,(\gotb-\gotB)]+\cot\beta\,\cos\theta}{\sin\theta}. 
\endeq
and the allowed $\varphi$-set  is $\underline{{\Phi}^C_{12}}$. 
Thus, we conclude that relation \refeq{5.3C} is obeyed within the set 
\begeq\label{5.28}
\underline{\Phi_{C}}\equiv
\begin{cases}
\underline{{\Phi^C}_{13}}\cap\underline{{\Phi}_{12}}\cap
\underline{\Phi_{24}}\cap\underline{\Phi_{II}} &\ \  \text {if}\  \gotb>\gotB,\\
\underline{{\Phi^C}_{13}}\cap\underline{{\Phi^C}_{12}}\cap
\underline{\Phi_{24}}\cap\underline{\Phi_{II}} &\ \ \text {if}\  \gotb<\gotB.
\end{cases}
\endeq  
Finally, we analyze the inequality present in the middle of \refeq{5.3D}, \ie\ 
$\oell\,<\,\ocLp$, referred to as the "1\,4"case. Proceeding as in 
the three previous cases, after putting 
\begeq\label{5.30}
\cos\gotf{14} \equiv \gotb/\sqrt{1+\gotb^2},\quad\quad 
\sin\gotf{14} \equiv 1/\sqrt{1+\gotb^2},
\endeq 
\begeq\label{5.31}
\cA_{14}\equiv \frac{\gota-\gotAp}{r\,\sqrt{1+\gotb^2}}\quad{\rm and}\quad 
\cB_{14}\equiv\frac {\gotb\,\cos\beta-\gotBp}{\sin\beta\,\sqrt{1+\gotb^2}},
\endeq 
we find that inequality $\oell\,<\,\ocLp$ becomes 
\begeq\label{5.33}
\sin(\varphi+\gotf{14})>\frac{\cA_{14}+\cB_{14}\,\cos\theta}{\sin\theta}
\endeq 
and the $\varphi$-range where it  is obeyed is 
\begeq\label{5.34}
\underline{\Phi_{14}}=[\phi_{14,L},\, \phi_{14,R}]
\endeq 
with 
\begin{eqnarray}\label{5.35}
\phi_{14,L}&\equiv&\gotf{14}+\arcsin\Big(
\frac{\cA_{14} +\cB_{14}\cos\theta}{\sin\theta}\Big),\\ 
\phi_{14,R}&\equiv& \pi +
\gotf{14}-\arcsin\Big(
\frac{\cA_{14} +\cB_{14}\cos\theta}{\sin\theta}\Big).\label{5.36}
\end{eqnarray}
Thus, the $\varphi$-range where relation \refeq{5.3d} holds true  is 
\begeq\label{5.37}
\underline{\Phi_{D}}\equiv\underline{{\Phi^C}_{13}}\cap\underline{{\Phi}_{14}}
\cap\underline{{\Phi^C}_{24}}\cap\underline{\Phi_{II}}. 
\endeq    
In this way we have determined the $\varphi$-domains, \ie\ \refeq{5.19a}, 
\refeq{5.19c}, \refeq{5.28} and \refeq{5.37}, where inequalities 
\refeq{3.14}, \refeq{3.15} and \refeq{3.20} can be fulfilled. In fact, one 
should still analyze if the relevant intersections do not yield void sets. 
This analysis, to be performed, requires the enumerations of all possible 
order relations among geometrical parameters:  $r$, $y_m$, $\gota$ 
and so on.  However the proof of the validity of property \bfPz\  does not 
require this further analysis. To this aim, it is sufficient to note that the 
end-points of domains $\underline{\Phi_{A}},\ldots,\underline{\Phi_{D}}$ 
necessarily are either of the form  
 $\gotf{}   +\arcsin[(\gotA+\gotB\cos\theta)/sin\theta]$ or are equal 
to constants as $\pm\pi/2$ or $3\pi/2$.  
\subsection*{6. Integral  \refeq{4.8} is an algebraic function}
We begin the proof of this point  by reporting first the expression of 
the integrand of \refeq{4.7}, for the cases $k=1,3$,  in terms of the new 
 variables 
\begeq\label{6.1}
t=\cos\theta\quad{\rm and}\quad u=\sin \phi.
\endeq
Functions $\cJ_1[r,\theta, \phi(\theta)]$ and  $\cJ_3[r,\theta, \phi(\theta)]$ 
respectively become
\begin{eqnarray}\label{6.2}
&&\gotF_1[r,\,t,\,u(t)] =\phi(t)\,\gotF_{1,A}(r,\,t)+ 
\gotF_{1,B}[(r,\,t,\,u(t)],\end{eqnarray}
with
\begin{eqnarray}
&&\gotF_{1,A}(r,t)\equiv -(\gota+\gotb\,r\,t\,\cot\beta)\,
t^2\,\cos\beta -(\gotb\,r\,t/2)\,(1-t^2)\,\sin\beta ,\label{6.3}\\
&&\gotF_{1,B}[r,t,u(t)]\equiv \,(\gotb\,r/2)\,\,t\,(1-t^2)\,u\,
\sqrt{1-u^2}\,\sin\beta -\nonumber\\
&&\quad\quad\quad\quad\quad\quad
t\,\sqrt{1-t^2}\,\sqrt{1-u^2}\,(\gota\,\sin\beta+2\,
\gotb\,r\,t\,\cos\beta),\label{6.4}
\end{eqnarray}
and 
\begin{eqnarray}\label{6.5}
&&\gotF_3[r,\,t,\,u(t )]=\phi(t)\,\gotF_{3,A}(r,\,t,)+ 
\gotF_{3,B}[r,\,t,\,u(t)],\end{eqnarray}
with
\begin{eqnarray}
&&\gotF_{3,A}(r,t)\equiv -t^2\,(\gotap\,\cos\beta+
\gotbp\,r\,t\,\cot\beta),\label{6.6}\\
&&\gotF_{3,B}[r,t,u(t)]\equiv  r\, t^2\,u\,\sqrt{1 - t^2}\,
\cos\beta-(r\,t/2)(1 - t^2)\,u^2\,\sin\beta
 -\nonumber\\
&&\quad\quad\quad\quad\quad\quad 
t\,\sqrt{1-t^2}\,\sqrt{1-u^2}\,(\gotap\,\sin\beta+
\gotbp\,r\,t).\label{6.7}
\end{eqnarray}
Functions $\gotF_2(r,\,t,\,u )$ and $\gotF_4(r,\,t,\,u )$, 
generated by $\cJ_2(r,\theta, \phi)$ and  
$\cJ_4(r,\theta, \phi)$, as well as their components  
$\gotF_{2,A}(r,\,t )$, $\gotF_{2,B}(r,\,t,\,u )$ and 
$\gotF_{4,A}(r,\,t )$, $\gotF_{4,B}(r,\,t,\,u )$, 
are respectively obtained from \refeq{6.2}-\refeq{6.4} and 
\refeq{6.5}-\refeq{6.7} by the parameter 
transformations reported in \refeq{4.3} and \refeq{4.5}. 
[All these $\gotF_{\dots}(\ldots )$ expressions  have 
been obtained setting $\cos\phi=\sqrt{1-u^2}$,  
implicitly assuming that $-\pi/2<\phi<\pi/2$. 
Whenever one had  $\pi/2<\phi<3\pi/2$ one should 
set  $\cos\phi=-\sqrt{1-u^2}$ but the following 
conclusions remain unchanged.]\\ 
The above $\gotF_{\dots}(\ldots )$ functions must be 
evaluated at each end-point $\phi(\theta)$ of the allowed 
$\varphi$-range and then integrated with respect to $t$ 
[note that $\sin\theta d\theta\, \to\, dt$]. 
Consider first the case where $\phi=const$. Then,  
$u(=\sin\phi)$ also is equal to a constant and all the 
$\gotF_j$s are polynomial functions of the only $t$ and 
$\sqrt{1-t^2}$. Consequently, their $t$-primitives are 
algebraic functions\expcite{18} of $t$. \\
We consider now to the more complicate case where the end 
points of the allowed $\varphi$-range are of  the  form 
\begeq\label{6.8m}
\phi= \gotf{}   +\arcsin\Big[
\frac{\gotA+\gotB\,\cos\theta}{sin\theta}\Big].
\endeq 
where $\gotf{}$,  $\gotA$ and $\gotB$ are suitable functions 
of the geometrical features of the outset two triangles as 
specified, \eg, in equations \refeq{5.5} and \refeq{5.8}.
In terms of variable $t=\cos\theta$, we find that 
\begeq\label{6.8}
\phi=\phi(t)=  \gotf{} +\arcsin\Big[
\frac{\gotA+\gotB\,t}{\sqrt{1-t^2}}\Big],
\endeq
where the use of the same symbol $\phi$ should not cause 
ambiguities. 
We find for $u=u(t)=\sin[\phi(t)]$ and 
$\sqrt{1-u^2}=\cos\phi(t)$ that
\begeq\label{6.9}
u= \frac{\sqrt{1-t^2-(\gotA+\gotB\,t)^2}}{\sqrt{1-t^2}}\,
\sin\gotf{}+  
\frac{\gotA+\gotB\,t}{\sqrt{1-t^2}}\,\cos\gotf{},
\endeq 
and 
\begeq\label{6.10}
\sqrt{1-u^2}= \frac{\sqrt{1-t^2-(\gotA+\gotB\,t)^2}}
{\sqrt{1-t^2}}\,\cos\gotf{}-   
\frac{\gotA+\gotB\,t}{\sqrt{1-t^2}}\,\sin\gotf{}. 
\endeq 
We see that the above two expressions are rational functions 
of  two radicals,  namely $\Delta\equiv\sqrt{1-t^2}$ and 
$\Delta_{1}\equiv\sqrt{1-t^2-(\gotA+\gotB\,t)^2}$.   
Consequently the substitution of \refeq{6.9} and \refeq{6.10} 
into the previous $\gotF_{j,B}$ ($j=1,\ldots,4$) definitions, that 
we shall first analyze, yields rational functions of $t$ and  
{\underbar {two}} radicals.   From this finding one would hastily 
conclude that  the $t$-primitives of the $\gotF_{j,B}$s 
are elliptical integral functions and not simple algebraic functions. 
However, a more careful consideration of the $\gotF_{j,B}$s 
shows that these involve variable $u$ in the forms:  $u^2$, 
$u\sqrt{1-u^2}$ and  $\sqrt{1-t^2}\sqrt{1-u^2}$. Thus,  
substituting  here equations \refeq{6.9} and \refeq{6.10} one 
obtains  expressions which only involve the radical $\Delta_1$. 
In fact,  the explicit substitution of 
\refeq{6.9} and \refeq{6.10} into \refeq{6.4} and \refeq{6.7} yields 
\begeq\label{6.11}
\gotF_{1,B}[r,t,u(t)]= \gotC_{1,B,1}(t) +\gotC_{1,B,2}(t)\,\Delta_{1}\\
\endeq
with 
\begin{eqnarray}
& & \gotC_{1,B,1}(t)\equiv (t/2) \sin \gotf{}\,  
 \Big[4\, \gotb\, r\, t (\gotA + \gotB\, t) \cos\beta + 
\Big(2 \gota (\gotA + \gotB\, t) -
\quad\quad \label{6.11a}\\ 
& &\quad\quad\quad\quad  \gotb\, r ( 2 \gotA^2 -1  + 4 
\gotA\, \gotB\, t + (1 + 2 \gotB^2) t^2) 
\cos \gotf{}\Big) \sin\beta\Big],\nonumber \\
& & \gotC_{1,B,2}(t)\equiv  (t/2) \Big [\gotb\, r (\gotA + 
\gotB\, t) \cos^2\gotf{} \,\sin\beta - 
   \gotb r (\gotA + \gotB\, t) \sin^2\gotf{} \sin\beta -
\quad  \label{6.11b} \\
& &\quad\quad\quad\quad   2 \cos\gotf{} (2 \gotb\, r\, t\, 
\cos\beta + \gota\, \sin\beta)\Big]\nonumber 
\end{eqnarray}
and 
\begeq\label{6.12}
\gotF_{3,B}[r,t,u(t)]= \gotC_{3,B,1}(t) +\gotC_{3,B,2}(t)\,\Delta_{1}
\endeq
with  
\begin{eqnarray} 
& & \gotC_{3,B,1}(t)\equiv (t/2) \Big[2\, r\, t (\gotA + 
\gotB\, t) \cos\gotf{}\, \cos\beta - 
   r\, (\gotA + \gotB\, t)^2 \cos\gotf{}^2 \sin\beta +
\nonumber \\    
 & &\quad\quad  \sin\gotf{} \Big(2 \gotbp\, r\, t 
(\gotA + \gotB t) + \big[2\, \gotap (\gotA + \gotB\, t) + 
  r\, (\gotA^2-1 + 2 \gotA\, \gotB\, t +\nonumber \\ 
&&\quad\quad  (1 + \gotB^2) t^2) \sin\gotf{}\big] 
\sin\beta\Big)\Big],\label{6.12a}\\
& & \gotC_{3,B,2}(t)\equiv -t \Big[-r\, t\, \cos\beta\, 
\sin\gotf{} + 
   \cos\gotf{} \Big(\gotbp\, r\, t\, +\nonumber \\
&& \quad\quad\quad\quad\quad\quad\sin\beta\,(\gotap + r\,
(\gotA + \gotB\, t) \sin\gotf{}\big) \Big)\Big].\label{6.12b}
\end{eqnarray}
The expressions of $\gotF_{2,B}(r,t,u(t))$ and 
$\gotF_{4,B}(r,t,u(t))$ are respectively obtained from 
the above two ones by the aforesaid transformations 
\refeq{4.3} and \refeq{4.5} of  coefficients 
$\gota, \gotb,\ldots$. One concludes that the $t$-primitives  of 
$\gotC_{j,B,1}(t)$, for $j=1,\ldots,4$, simply are  $t$-polynomials 
of the 4th-degree. The primitives of   $\gotC_{j,B,2}(t)\Delta_1$ 
(for $j=1,\ldots,4$) are obtained by the procedure expounded in  
Ref. [18]. 
For definiteness we take $j=1$ and we observe 
that $-(1+\gotB^2)$, the coefficient of $t^2$ inside radical $\Delta_1$, 
is negative. Since the integral must be real, variable $t$ must vary 
between the lowest ($\mu_1$) and the largest root ($\mu_2$) of 
the equation $1-t^2-(\gotA+\gotB\,t)^2=0$. We change 
the integration variable $t$ into the new variable $\xi$ according to 
\begeq\label{A.1}
t =t(\xi)\equiv  \frac{\mu_2+\mu_1 \xi^2}{ 1+\xi^2}. 
\endeq 
the inverse of which is $\xi=\sqrt{(\mu_2-t)/(t-\mu_1)}= 
\frac{\Delta_1}{(t-\mu_1)\sqrt{1+\gotB^2}}$.
We find that 
\begeq\label{A.2}
\Delta_1 = \frac{(\mu_2-\mu_1)\,\xi}{1+\xi^2}\quad{\rm and}\quad  
\Big\vert\frac{d\,t}{d\xi}\Big\vert=
\frac{2(\mu_2-\mu_1)\xi}{(1+\xi^2)^2}. 
\endeq
The  new $\xi$-integrand, also 
accounting for the Jacobian, is
\begeq\nonumber  
2\,\sqrt{1+\gotB^2}\, \gotC_{1,B,2}[t(\xi)]\,
\frac{(\mu_2-\mu_1)^2\,\xi^2}{(1+\xi^2)^4} 
\endeq 
that is a rational function of $\xi^2$. The relevant $\xi$-primitive 
is a rational function of $\xi$ plus a contribution proportional to 
$\arctan\,\xi$. Hence, the primitive of $\gotC_{j,B,2}(t)\Delta_1$ 
is a rational function of $t$ and $\Delta_1$ plus an inverse 
trigonometric contribution  function of the same two variables. \\ 
To complete the proof of \bfPz\ for the case of non-parallel 
facets, we must show that the other contributions of 
the form $\arcsin[u(t)]\,\gotF_{j,A}(r,\,t)$  also are 
integrable in algebraic form.    For definiteness we again 
consider the case $j=1$. 
An integration by parts of $\arcsin[u(t)]\,\gotF_{1,A}(r,\,t)$ 
yields
\begeq\label{6.13}
 \gotP_{1,A}(r,\,t)\,\arcsin[u(t)] - \int \gotP_{1,A}(r,\,t)
\Big(\frac{d\, \arcsin[u(t)]}{dt}\Big)dt
\endeq 
where  
\begeq\label{6.14}
\gotP_{1,A}(r,\,t)\,\equiv\, \int \gotF_{1,A}(r,\,t) dt,
\endeq  
certainly is an algebraic  $t$-function since it  
is  a 4th-degree $t$-polynomial. 
The $\arcsin[u(t)]$ derivative reads  
\begeq\label{6.15}
\frac{d\, \arcsin[u(t)]}{dt}= 
\frac{\gotB+\gotA\,t}{(1-t^2)\,\Delta_1}.
\endeq
It  involves the only  radical $\Delta_1$. Then, the integrand 
present in \refeq{6.13}  is a rational function of $t$ and of 
$\Delta_1$ and  its primitive can explicitly be 
evaluated as reported in the first part of the appendix.   
In this way we have completed the proof that each of 
the integrals present in equation \refeq{4.8} has an 
algebraic form and the proof of \bfPz\ 
for the case of polyhedrons with no pair of parallel 
facets is achieved.  
\subsection*{7. The case of the parallel facets}
To complete the proof of \bfPz\  we must show that, 
whenever the considered polyhedron has parallel facets,  
the contributions of these to the 
CLD also are algebraic functions. \\ 
\begin{figure}[h]
 {\includegraphics[width=10.truecm]{{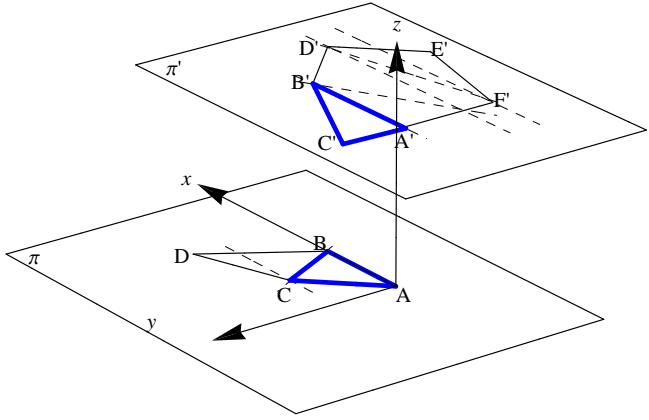}}}
\caption{\label{FigA2} { ABCD   and  A'C'B'D'E'F' 
represent the polyhedron's facets that lie 
on  parallel planes $\pi$ and $\pi'$.  The figure 
shows how to decompose the two facets into 
a set of triangles.   }}
\end{figure}
With reference to Fig.\,2, let  ABCD   and  A'C'B'D'E'F' be  
the parallel facets of the given polyhedron.  Denote them 
by $\cS_1$ and $\cS_2$ and  the planes where these lie 
by $\pi$ and $\pi\p$. 
We choose one of their sides as the direction for 
carrying out the "triangularization" of the facets. 
In Fig.\,2 we have chosen side AB.  Through  each vertex 
we draw a line parallel to AB till 
intersecting the opposite side and then the resulting 
trapezia are bisected by one of their 
diagonals. (The cutting lines are  broken  in Fig.\,2.)  
By so doing,  ABCD is decomposed into three triangles 
and A'C'B'D'E'F' into six. Contribution \refeq{2.4}, 
due to the present $\cS_1$ and $\cS_2$, is equal to  
the sum of  contributions of the form \refeq{2.5} 
where $\cT$ and $\cT\p$ respectively span all the 
triangles of $\cS_1$ and of $\cS_2$. Consider, for 
definiteness, the contribution where $\cT$=ABC and 
$\cT\p$=A'B'C'. We choose the right-handed cartesian 
frame $Oxyz$ by taking axis $x$ along AB, axis $y$ 
lying onto $\pi$ and orthogonal to $x$  
and axis $z$ orthogonal to $\pi$ and $\pi\p$. Besides, 
eventually renaming some vertices of 
the triangles, we can always have the case depicted in 
the figure where the origin  coincides with A and 
the $y$-ordinate of C  as well as the $z$-ordinate of 
$\pi\p$ are both positive. Furthermore, 
we can always assume that $\hnu_1=\hnu_2=(0,0,1)$ by 
eventually changing the sign of integral \refeq{2.5}.  
We denote by $h$ the distance between the two parallel 
planes and by $\br_1=(x,y,0)$  and 
$\br_2=(X,Y,h)$ the coordinates of a generic point of 
$\cT$ and of  $\cT\p$, respectively.  
We also agree   that sides AC,  BC have equations
\begeq\label{7.1}
\ell(y)=\gota+\gotb\,y\quad{\rm and}\quad 
\cL(y)=\gotA+\gotB\,y
\endeq
and sides A'C' and B'C' 
\begeq\label{7.2}
\ellp(Y)=\gotap+\gotbp\,Y\quad{\rm and}\quad 
\cLp(Y)=\gotAp+\gotBp\,Y.
\endeq
Then, for triangles $\cT$ and $\cT\p$,  we respectively  
have that:
\begin{eqnarray}\label{7.3}
\ell(y) < \cL(y)&& \quad  {\rm if}\quad 0<y<y_M;\\
\ellp(Y) < \cLp(Y)&&\quad   {\rm if}\quad 
Y_m<Y<Y_M.\label{7.4}
\end{eqnarray}
Equation \refeq{2.5} 
takes now the form
\begin{eqnarray}\label{7.4}
&&g_p(r)= 
-\frac{1}{4\pi V}\int d\varphi\int \cos^2\theta\,
\sin\theta\,d\theta
\int_{0} ^{y_M}dy\int_{\ell(y)}^{\ocL(y)} dx \times\quad\\
&&\quad\quad\quad\int_{Y_m} ^{Y_M}dY
\int_{\ellp(Y)}^{\cLp(Y)} dX \cos^2\theta
\delta(\br_1+r\hw-\br_2),\nonumber
\end{eqnarray}
while  the Dirac function yields the equations 
\begin{eqnarray}
&&x+r\,\sin\theta\,\cos\varphi-X=0,\label{7.4a}\\
&&y+r\,\sin\theta\,\sin\varphi-Y=0,\label{7.4b}\\
&& r\,\cos\theta-h=0.\label{7.4c}
\end{eqnarray} 
[Suffix $p$ on the left  side of \refeq{7.4} 
recall us that the integral refers to parallel  
facets. Polar coordinates $(\theta,\varphi)$ of $\hw$ 
have been defined as in section\,3.]   
We solve with respect to $X$, $Y$ and  $\theta$. The 
solutions are: 
\begin{eqnarray}
{\bar\theta}&=& \arcos(h/r),\label{7.5a}\\
{\oX}={\oX}(x)&\equiv&x+\sqrt{r^2-h^2}\,\cos\varphi,\label{7.5b}\\
{\oY}={\oY}(y)&\equiv&y+\sqrt{r^2-h^2}\,\sin\varphi.\label{7.5c}
\end{eqnarray}
Performing the integrations with respect to 
$\theta$, $X$ and $Y$, integral \refeq{7.4} 
becomes
\begeq\label{7.6}
g_p(r)=-\frac{h^2}{4\pi\,r^3\, V}\int d\varphi
\int_{0} ^{y_M}dy\int_{\ell(y)}^{\cL(y)} \Theta(\cdot)dx,  
\endeq
where $\Theta(\cdot)=1$ if: $r>h$, $Y_m<{\oY}(y)<Y_M$ and  
$ \ellp({\oY}) <{\oX} <\cLp({\oY})$,  and equal to zero 
elsewhere. 
Once we substitute, in the last inequality,  ${\bar X}$ 
with expression \refeq{7.5b} we 
obtain a further inequality for $x$ beside the one specified  
in the  integral over $x$. 
The combination of these two inequalities yields  
\begeq\label{7.7}
\max[\ell(y), \,\oellp(y)] \,<\,x\,<\,
\min[\cL(y),\,\ocLp(y)]
\endeq
where we have put 
\begin{eqnarray}
&& \oellp({y})\equiv \ellp({\oY}(y))\,-
\sqrt{r^2-h^2}\cos\varphi,\label{7.8a}\\
&& \ocLp({y})\equiv \cLp({\oY}(y))-
\sqrt{r^2-h^2}\cos\varphi.\label{7.8b}
\end{eqnarray}
Owing to \refeq{7.7} the integral over $x$ can 
be performed  and yields
\begeq\label{7.9}
g_p(r)=-\frac{h^2}{4\pi\,r^3\, V}\int d\varphi
\int_{0} ^{y_M}\Big[\min[\cL(y),\,\ocLp(y)]-
\max[\ell(y), \,\oellp(y)]\Big]\,dy. 
\endeq
This integral is similar to  \refeq{3.23}. Thus, we  
apply a procedure similar to that described 
in sections 4-6 in order to show that $g_p(r)$ also 
has an algebraic form. In this case 
the analysis is somewhat simpler. First, we observe 
that the  integrand of \refeq{7.9} involves the same 
four cases reported in Eq.s \refeq{5.3A}-\refeq{5.3D} 
and its value is respectively equal to $\ocL(y)-\oellp(y)$, 
$\ocLp(y)-\oellp(y)$, $\ocL(y)-\oell(y)$ and $\ocLp(y)-\oell(y)$. 
Thus, the integrand  always is a linear function of $y$,  
$\cos\varphi$ and $\sin\varphi$. 
We must also require that $Y_m<\oY(y)<Y_M$ in agreement with 
\refeq{7.4}. This inequality 
combined with  $0<y<y_M$ yields
\begeq\label{7.11}
\max[0,\, Y_m-\sqrt{r^2-h^2}\sin\varphi]\,<\,y\,<\,
\min[y_M,\,Y_m-\sqrt{r^2-h^2}\sin\varphi].
\endeq
The reduction of the inequalities, related to the left and right 
sides  of \refeq{7.11}, bounds the $\varphi$-range,  the 
end-points of which are of the form 
\begeq\label{7.11a}
\varphi< f_1+\arcsin[R_1(\sqrt{r^2-h^2},\ldots)]\ {\rm or}\ 
\varphi> f_2+\arcsin[R_2(\sqrt{r^2-h^2},\ldots)],
\endeq 
where $f_1$ and $f_2$ denote suitable constants,  $R_1$ and $R_2$ 
rational functions and the dots 
other geometrical parameters as $y_m,\ldots,Y_M$. 
The remaining $y$-inequalities, associated to  \refeq{7.11}  and 
to \refeq{5.3A}-\refeq{5.3D}, determine the corresponding intervals 
of the accceptable $y$-values. Since they are linear in $y$ it follows 
that  the left and right bound of each interval are of the form  
\begeq\label{7.13}
y_B= \gotC(\ldots)+\gotD(\ldots)\,\sin\varphi,
\endeq
where $\gotC$ and $\gotD$ are functions of 
some of the geometrical parameters. generically denoted 
by the dots in \refeq{7.13}.  We remark that $y_B$ is a 
linear function of $\sin\varphi$. Thus, in  taking the 
intersection of the previous $y$-intervals, we need to 
compare the $y$-bounds among themselves. This 
comparison restricts the $\varphi$-range but we are 
certain that the resulting $\varphi$-bounds are of 
the form $f+\arcsin[R(\sqrt{r^2-h^2},\ldots)]$, 
where $R$ again denotes a rational functions and the  
dots the involved geometrical parameters. We can now proceed 
to establish the functional form of integral \refeq{7.9}. 
We already noted that the integrand always is a liner 
$y$-function. Thus, its primitive is a 2nd-degree 
$y$-polynomial. Once this is evaluated at the 
ends of the $y$-integration domain, these points 
are of the form \refeq{7.13}. Consequently, the 
primitive value is a 2nd-degree polynomial of 
$\sin\varphi$ and the resulting $\varphi$-primitive 
will be a 2nd-degree polynomial of $\sin\varphi$ and 
$\cos\varphi$ plus a term linear in $\varphi$. 
[The explicit expression is reported at the end  
of the appendix.]  
This proves that $g_p(r)$ also has an algebraic form and 
statement {\bfPz} is now fully proved. 
\subsection*{8. Conclusions} 
We have shown that the CLD of any polyhedron is a sum of 
integrals of the form  \refeq{2.4} that, in turns, converts 
into  a  sum of  contributions of  the forms \refeq{3.23} 
and \refeq{7.6} for each pair of non-parallel and parallel 
facets, respectively. After having reduced the corresponding 
inequalities with respect to the variable of the innermost 
integrals, we have shown that the final primitives, with 
respect to  $t$ [for definiteness we consider the case of the 
non-parallel facets, but the results also apply to the 
parallel ones],  always are a sum of a rational functions 
and an inverse trigonometrical function, the argument of 
which is a rational function of $t$  and the square root of  
a 2nd-degree polynomial of $t$. The primitives have to be 
evaluated at the end points of the allowed range of $t$. 
These bounds  depend on $r$ and the 
parameters that describe  the geometry of the facets. 
Even though we have not fully determined their explicit 
expressions, we have shown that they 
are of the form $\arcsin[R(r,\dots)]$, $R[\cdot]$ being 
a rational function and the dots denoting the appropriate 
geometrical parameters.  The presence of the $arcsin$ 
functions both in the bounds and in the primitive implies 
that the (higher order) $r$-derivatives of the final CLD 
can only show singularities of the form 
$|r-\delta_j|^{-n}$  and $|r-\delta_j|^{-m+1/2}$ with $n$ 
and $m$ positive integers and $\delta_j$ depending on 
the geometrical parameters. 
The first kind of singularity follows from the rational 
function contributions and the second from the 
inverse trigonometric functions as well as from the fact 
that $R$ depends on the square root of a polynomial.  
Finally,  on physical grounds, it can be stated that each 
of the above $\delta_j$s must be  one of  end-points 
of the different $r$-subdomains where the CLD has a 
specific functional form.  Hence, 
the following general property:\\
{\bf $\bf {P_3}$} - {\em  the derivatives of the CLD of 
any polyhedron, 
as $r$ approaches one of the end points (say $\delta_j$) of 
the $r$-subdomains, can 
only show algebraic singularities of   form  
$(r-\delta_j)^{-n}$ or $(r-\delta_j)^{-m+1/2}$ with n and m 
positive integers.} \\ 
Illustrations of this property can easily be found by looking 
at the derivatives of the algebraic known CLDs of the first three 
Platonic solids that we mentioned earlier. Besides {\bf $\bf {P_3}$} 
is useful in deriving the sub-leading terms of the asymptotic 
expansion of the Fourier transform of the CLD  (see, 
\eg, Ref. [19]). 
\vfill\eject
\subsection*{Appendix: Primitives' explicit expressions }
For completeness we report below the explicit expressions of the $t$-primitives. 
\subsubsection*{The non parallel case}
We consider first the case of the non-parallel facets. We  recall that,  after having 
reduced all the inequalities, in each of the resulting cases we must perform an 
integration over variable $\theta$ of an integrand which, according  to equation 
\refeq{4.6}, is the algebraic sum of four contributions of the form \refeq{4.8} where, 
as shown in section 5, the most general form of $\phi_B(\theta)$ is: 
$\gotf{}+\arcsin[\frac{\gotA+\gotB\,\cos\theta}{\sin\theta}]$. Passing to the 
new integration variable $t=\cos\theta$, the integrand functions become the 
functions $\gotF_1[r,\,t,\,u(t) ]$, $\gotF_2[r,\,t,\,u(t) ]$, $\gotF_3[r,\,t,\,u(t)]$ 
and $\gotF_4[r,\,t,\,u(t)]$. The first and the third are respectively defined 
by equations  \refeq{6.2} and \refeq{6.5}, with $u(t)=\sin\phi(t)$ and 
$\phi(t)=\gotf{}+\arcsin\Big[\frac{\gotA+t\,\gotB}{\sqrt{1-t^2}}]$ [see \refeq{6.1} 
and \refeq{6.8}], while the remaining two are obtained from the former ones by 
applying the appropriate substitutional relation. According to equations \refeq{6.2} 
and \refeq{6.5}, both $\gotF_1[r,\,t,\,u(t)]$ and  $\gotF_3[r,\,t,\,u(t)]$ are  sums of 
two functions characterized by the further index$A$ or $B$. The explicit calculations, 
carried out with the help of the MATHEMATICA software  along the lines expounded 
in section 6, show that the primitive of $\big[\phi(t)\,\gotF_{1,A}(r,\,t)]\big]$ is
 equal to 
\begeq\label{a.1}
\gotP_{1,A,a}(t)+\gotP_{1,A,b,I}(t)+\gotP_{1,A,b,0}( t)+\gotP_{1,A,b,1}( t)+
\gotP_{1,A,b,2}( t)+\gotP_{1,A,b,3}( t), 
\endeq
and that of $\gotF_{1,B}[r,\,t,\,u(t)]$ to
\begeq\label{a.2}
\gotP_{1,B,a}(t)+\gotP_{1,B,A}(t)+\gotP_{1,B,B}( t). \endeq 
Similarly, the primitive of $\big[\arcsin[u(t)]\,\gotF_{3,A}(r,\,t)\big]$ is given by 
\begeq\label{a.3} 
\gotP_{3,A,a}(t)+\gotP_{3,A,b,I}(t)+\gotP_{3,A,b,0}( t)+\gotP_{3,A,1}( t)+
\gotP_{3,A,2}( t)+\gotP_{3,A,3}( t), \endeq   
and that of $\gotF_{3,B}[r,\,t,\,u(t)]$ by 
\begeq\label{a.4}
\gotP_{3,B,a}(t)+\gotP_{3,B,A}(t)+\gotP_{3,B,B}( t). 
\endeq 
All the above $\gotP_{\dots}( t)$ functions are reported below.\\
The primitives of $\gotF_2[r,\,t,\,u(t)]$ and $\gotF_4[r,\,t,\,u(t)]$ are respectively  
obtained from that of $\gotF_1[r,\,t,\,u(t)]$ and that of $\gotF_3[r,\,t,\,u(t)]$ by 
the appropriate substitutional rule. It is also noted that the sign ambiguity reported 
below equation \refeq{6.7} does not invalidate the derivation of the above 
$t$-primitives. In fact, whenever we had to put $\cos\varphi=-\sqrt{1-u^2}$ into 
equation \refeq{6.4} [or \refeq{6.7}], it is sufficient to change the signs of $\gota$ 
and $\gotb$ [$\gotap$ and $\gotbp$] to recover the integrand form to which 
\refeq{a.2} [\refeq{a.4}] applies. \\
The primitive expressions, with the definitions of the related quantities, read as follows:
\begin{eqnarray}\label{A.0}
&& \Delta\equiv\sqrt{1-t^2},\quad \Delta_1\equiv\sqrt{1-t^2-(\gotA+\gotB\,t)^2},\\
&& \Lambda_1\equiv\sqrt{1+\gotB^2},\quad\quad \Lambda_2\equiv\sqrt{1-\gotA^2+\gotB^2},
\end{eqnarray}
\begin{eqnarray}\label{A.1}
&&c_{1,A,a,1} =  -(1/4) \gotb\,  r\, \sin\beta,\quad 
c_{1,A,a,2} = -(1/3) \gota\,  \cos\beta,\\
&& c_{1,A,a,3} = -(1/16)
     \gotb \, r\, [1 + 3 \cos(2 \beta)]\, \csc\beta,\\
&&\gotP_{1,A,a}(t)=  \gotf\,\sum_{j-1}^3\,c_{1,A,a,j} t^{3-j} 
\end{eqnarray}
\begin{eqnarray}\label{A.4}
&&c_{1,A,b,I,1} = -(1/4) \gotb\, r \sin(\beta),\quad 
c_{1,A,b,I,2} = -(1/3)\gota\, \cos(\beta),\\
&&c_{1,A,b,I,3} = -(1/16)\gotb\, r [1 + 3 \cos(2 \beta)] \csc\beta,\\
&&\gotP_{1,A,b,I}( t)=
 \arcsin\big[\frac{\gotA + \gotB\, t}{\Delta}\big]\, \sum_{j=1}^3 c_{1,A,b,I,j}\, t^{j+1},
\end{eqnarray}
\begin{eqnarray}\label{A.8}
& &c_{1,A,b,0,1,1}=8\, \gota\, \gotB\, \Lambda_1^2 [2 + 2\, \gotB^2 - 3\, \Lambda_1^2 + 
3\, \Lambda_2^2]\, \cos\beta, \\
& &c_{1,A,b,0,1,2}= \gotA\, \gotb\,\{ 2\, [8\, \Lambda_1^4 + 15\, \Lambda_2^2 - 
\Lambda_1^2\, (2 + 11\, \Lambda_2^2) ]\,\csc\beta + \\
& &\quad\quad \quad\quad\quad\quad     
3\, [ \Lambda_1^2\, (2 + 11\, \Lambda_2^2)-4\, \Lambda_1^4 - 15\, \Lambda_2^2] \sin\beta\},\nonumber\\
& & c_{1,A,b,0,2,1}=8\, \gota\, \gotA \,\Lambda_1^4\, \cos\beta, \\
& &c_{1,A,b,0,2,2}=\gotb\,\gotB\,\Lambda_1^2\,(5\,\Lambda_2^2 - 2\,\Lambda_1^2)\,
(2\, \csc\beta - 3\, \sin \beta),\\
& &c_{1,A,b,0,3,1}=0,\quad  c_{1,A,b,0,3,2}= 
   \gotA\, \gotb\, \Lambda_1^4 [1 + 3\, \cos(2\, \beta)]\, \csc \beta,\\
&&\gotP_{1,A,b,0}( t)=\frac{\Delta_1}{48 \Lambda_1^6}\,
\sum_{i=1}^3\sum_{j=1}^2 c_{1,A,b,0,i,j}\,t^{i-1}\,r^{j-1};
\end{eqnarray}
\begin{eqnarray}\label{A.12}
& &\Omega_{1,A,b,1}=\frac{\Lambda_1\, \Delta_1}{
 \gotA\, \gotB + \Lambda_2 + \Lambda_1^2\, t},\\
& & c_{1,A,b,1,1}= 16\, \gota\, \gotA \,[3 -\gotA^2 + 
(3 + 2 \gotA^2)\, \gotB^2]\, \Lambda_1^2\, \cos\beta, \\
&&c_{1,A,b,1,2}=  3\, \gotb\, \gotB (\Lambda_1^2-\gotA^2) \{7 - 3\, \gotA^2 + 
(13 + 2\, \gotA^2)\, \gotB^2 +6\, \gotB^4 +\nonumber \\
&& \quad\quad\quad\quad   [5 - 9\, \gotA^2 + (7 + 6\, \gotA^2)\, \gotB^2 + 
2\, \gotB^4] \,\cos(2\, \beta)\} \,\csc\beta, \\
 &&\gotP_{1,A,b,1}( t)=\frac{\arctan(\Omega_{1,A,b,1})}  {48\, \Lambda_1^7}\,
\sum_{j=1}^2\,c_{1,A,b,1,j} \,r^{j-1} ;
\end{eqnarray}
\begin{eqnarray}
&&\Omega_{1,A,b,2}=\frac{(\gotA + \gotB) \Delta_1}
{1 - \gotA\, \gotB - \gotA^2 - \Lambda_2 - 
(1 + \gotA\, \gotB + \gotB^2 - \Lambda_2)\, t},\\
&& \gotP_{1,A,b,2}( t)=\frac{\arctan(\Omega_{1,A,b,2})\,\csc\beta}{48 }\times \\
&&\quad\quad\quad \quad\quad\quad 
\{8\, \gota\, \sin(2\,\beta) + 3\,\gotb\, r\, [3 + \cos(2\, \beta)]\}\,\nonumber;
\end{eqnarray}
\begin{eqnarray}
&&\Omega_{1,A,b,3}= \frac{(\gotA - \gotB)\, \Delta_1}{1 - \gotA^2 + 
 \gotA\, \gotB + \Lambda_2 + (1 - \gotA\, \gotB + \gotB^2 + \Lambda_2)\, t},\\
&& \gotP_{1,A,b,3}( t)= -\frac{\arctan(\Omega_{1,A,b,3})\, \csc\beta}{48}\times\\
 &&\quad\quad\quad\quad\quad \quad\quad  
\{8\, \gota\, \sin(2\, \beta) - 3\, \gotb\, r\, [3 + \cos(2\, \beta)]\};\nonumber
\end{eqnarray}                                  
\begin{eqnarray}
&&c_{1,B,a,1}=  \frac{\sin\gotf{}\, \sin\beta}{4} [2\, \gota\, \gotA\, + (1 - 2\, \gotA^2) \gotb\, r \cos\gotf{}] , \\
&&c_{1,B,a,2}=\frac{\sin\gotf{}}{3}\, [2\, \gotA\, \gotb\, r\, \cos\beta + 
     \gotB\, (\gota\, - 2\, \gotA\, \gotb\, r\, \cos\gotf{})\, \sin\beta)] ,\\
&&c_{1,B,a,3}= -\frac {\gotb\, r\, \sin\gotf{}}{8}\, [(1 + 2\, \gotB^2)\, \cos\gotf{}\, \sin\beta - 
4\, \gotB\, \cos\beta],\\
&& \gotP_{1,B,a}( t)= \,\sum_{j=1}^3\,c_{1,B,a,j} \, t^{j+1}\,;
\end{eqnarray}        
\begin{eqnarray}                                 
&& c_{1,B,A,1,1}=8\, \gota\, [2 - 2\, \gotA^2 + (2 + \gotA^2)\, \gotB^2]\, \Lambda_1^2 
\cos\gotf{}\, \sin\beta\,  , \\
&& c_{1,B,A,1,2} = -\gotA\, \gotb\, \{4\, \gotB\, [\gotA^2\, ( 2\, \gotB^2-13 ) + 13\, \Lambda_1^2]\, \cos\gotf{}\, \cos\beta + \\
&&\quad\quad\quad [8 + 3\, \gotB^2 - 5\, \gotB^4 + 
          \gotA^2 ( 9\, \gotB^2 + 2\, \gotB^4-8 )]\,\sin\beta\, \cos(2 \gotf{})\},\nonumber\\
&& c_{1,B,A,2,1}= -8\,\gota\, \gotA\, \gotB\, \Lambda_1^4 
       \cos\gotf{} \sin\beta, \\
&& c_{1,B,A,2,2} = \gotb\, \Lambda_1^2 \{4\, [\gotA^2 ( 2\, \gotB^2-3 ) + 3\,\Lambda_1^2]\, 
\cos\gotf{}\, \cos\beta + \\
&&\quad\quad\quad\quad\quad \gotB\, [\gotA^2\, (7 + 2\, \gotB^2) - 3\, \Lambda_1^2]\,
\cos(2 \gotf{})\, \sin\beta\}, \nonumber \\
&& c_{1,B,A,3,1}=-16\,\, \gota\, \Lambda_1^6\, \cos\gotf{}\, \sin\beta, \\
&& c_{1,B,A,3,2}=2\, \gotA\, \gotb\, \Lambda_1^4\, 
[ (4 + 5\, \gotB^2)\, \cos(2 \gotf{})\,\sin\beta\,-4\, \gotB\, \cos\gotf{}\, \cos\beta ],\\
&& c_{1,B,A,4,1}=0,\quad\\
&& c_{1,B,A,4,2}=6\, \gotb\, \Lambda_1^6\, [\gotB\, \cos(2 \gotf{})\, \sin\beta-
4\, \cos\gotf{}\, \cos\beta],\\
&& \gotP_{1,B,A}( t)=  \frac{\Delta_1}{48\, \Lambda_1^6}\,
   \sum_{i=1}^4\sum_{j=1}^2\,c_{1,B,A,i,j}\,t^{i - 1}\, r^{j - 1}\,;
\end{eqnarray}
\begin{eqnarray}
&& c_{1,B,B,1}=-8\, \gota\, \gotA\, \gotB\, \Lambda_1^2\, \cos\gotf{}\, \sin\beta, \\
&& c_{1,B,B,2}=   4\, \gotb\, [4\, \Lambda_1^4 + 5\, \Lambda_2^2 - 
       4\, \Lambda_1^2\, (1 + \Lambda_2^2)] \cos\gotf{}\,\cos\beta + \\
&&\quad\quad\quad\quad\quad     \gotb\, \gotB (4 \Lambda_1^2 - 5 \Lambda_2^2)\, 
\cos(2 \gotf{})\, \sin\beta,\nonumber\\
&&\gotP_{1,B,B}( t)= \frac{\Lambda_2^2}{8\, \Lambda_1^7}\, 
    \arctan\Big[\frac{\Delta_1\, \Lambda_1}
    {\gotA\, \gotB  + \Lambda_2+ t\, \Lambda_1^2}\Big] 
\sum_{j=1}^2\, c_{1,B,B,j}\,r^{j-1}\,;\\
&& \nonumber      \\
&&    \nonumber  
\end{eqnarray} 
\begeq                                              
\gotP_{3,A,a}( t)=-\frac{\gotf{} \,t^3\,\cos\beta }{12}\, 
\Big(4\,\gotap\,  + \frac{3\, \gotbp\,\, r\, t}{\sin\beta}\Big);
\endeq
\begin{eqnarray}
&&\gotP_{3,A,b,I}(t)= -\frac{t^3\, \cos\beta}{12} 
\arcsin\Big[\frac{\gotA + \gotB\, t}{\Delta}\Big]\, 
\Big(4\, \gotap\,  + \frac{3\, \gotbp\,\, r\, t}{\sin\beta}\Big ) ;
\end{eqnarray}
\begin{eqnarray}
&&c_{3,A,b,1}= \gotA\, ( 8\, \Lambda_1^4 -2\, \Lambda_1^2 + 
     15\, \Lambda_2^2 - 11\, \Lambda_1^2\, \Lambda_2^2),\\
&&c_{3,A,b,2}= -\gotB\, \Lambda_1^2\, (2\, \Lambda_1^2 - 5\, \Lambda_2^2),\quad
c_{3,A,b,3}= 2\, \gotA\, \Lambda_1^4,\\ 
&&\gotP_{3,A,b,0}(t)=\frac{\Delta_1\,  \cos\beta}{24\, \Lambda_1^6}  
\Big(\frac{r\, \gotbp\,\ \sum_{j=1}^3\,c_{3,A,b,j}\ t^{j-1} } {\sin\beta} +\\
&& \quad\quad\quad\quad\quad\quad  
    4\, \gotap\,\,  \Lambda_1^2\, [\gotB\, (2\, \Lambda_1^2 - 3\, \gotA^2) + 
       \gotA\, \Lambda_1^2\, t] \Big)\nonumber;
\end{eqnarray}
\begin{eqnarray}
&& \Omega_{3,A,1}=  \frac{\Delta_1\, \Lambda_1}
{ \gotA\, \gotB + \Lambda_2+ t\, \Lambda_1^2 }, \\  
&&c_{3,A,1} = 
 4\, \gotA\,\gotap\,\, \Lambda_1^2\, (2\, \Lambda_1^4 + 3\, \Lambda_2^2 - 
    2\, \Lambda_1^2\, \Lambda_2^2),\\
&&c_{3,A,2}  = 3\, \gotB\, \gotbp\,\, r\, \Lambda_2^2\, [4\, \Lambda_1^4 + 
    5\, \Lambda_2^2 - 2\, \Lambda_1^2\, (2 + \Lambda_2^2)],\\ 
&&\gotP_{3,A,1}(t)= \frac{ \arctan[\Omega_{3,A,1}]\,\cos\beta\,}{12\, \Lambda_1^7}
   \,\sum_{j=1}^2\,c_{3,A,j} \, \sin^{1-j}\beta;
\end{eqnarray}
\begin{eqnarray}
&& \Omega_{3,A,2}= \frac{(\gotA + \gotB)\, \Delta_1}
{\gotA\, \gotB + \Lambda_1^2 + \Lambda_2 - \Lambda_2^2 - 1 +  
t\, (\gotA\, \gotB + \Lambda_1^2 - \Lambda_2)},\\ 
&&\gotP_{3,A,2}(t)= -\frac{ \arctan[\Omega_{3,A,2}]\,\cos\beta }{12} \,
\Big[4\, \gotap\, + \frac{3\, \gotbp\,\, r}{\sin\beta}\Big];
\end{eqnarray}                                          
\begin{eqnarray}
&& \Omega_{3,A,3}=\frac{(\gotA - \gotB) \Delta_1}{(1 + 
    \gotA\, \gotB - \Lambda_1^2 + \Lambda_2 + \Lambda_2^2 + 
    t (\Lambda_1^2 + \Lambda_2 - \gotA\, \gotB)},\\ 
&&\gotP_{3,A,3}(t)= -\frac{\arctan[\Omega_{3,A,3}]\,\cos\beta}{12} \, 
\Big[4\, \gotap\, - \frac{3\, \gotbp\,\, r}{\sin\beta}\Big];
\end{eqnarray}
\begin{eqnarray}
&&c_{3,B,a,1}  =  \frac{\sin\beta}{8}\, \{4\,\gotA\,\gotap\,\,\sin\gotf{} - 
r\, [1 - \cos(2 \gotf{}) + 2\, \gotA^2 \cos(2 \gotf{})] \},\\
&&c_{3,B,a,2}  =  \{ 
 r\, \gotA\,[\cos\gotf{}\,\cos\beta + \gotbp\, \sin\gotf{} - 
         \gotB\, \cos(2 \gotf{})\, \sin\beta ]+\\
&&\quad\quad\quad\quad\gotap\,\,\gotB\,\sin\gotf{}\, \sin\beta \}/3,\nonumber\\ 
&&c_{3,B,a,3}  =
 \frac{r}{16} \big\{4\, \gotB\, \cos\gotf{}\, \cos\beta + 4\, \gotB\, \gotbp\,\, \sin\gotf{} + \\
&&\quad\quad\quad\quad  
    \sin\beta\, [1 - (1 + 2\, \gotB^2) \cos(2 \gotf{}) ]\big\},\nonumber\\
&&\gotP_{3,B,a}(t)=  \,\sum_{j=1}^3\, c_{3,B,a,j}\, t^{j+1};
\end{eqnarray}
\begin{eqnarray}
&&c_{3,B,A,1,1}=4\, \gotap\, \Lambda_1^2\, [\gotA^2\, (\gotB^2-2) + 
      2\, \Lambda_1^2]\, \cos\gotf{} \,\sin\beta, \\      
&&c_{3,B,A,1,2}= \gotA\,\gotB\,[\gotA^2\, ( 2\, \gotB^2-13 ) + 13\, \Lambda_1^2]\, 
( \cos\beta \sin\gotf{}-\gotbp\, \cos\gotf{}) +\quad\quad\\
&& \quad\quad\quad\quad\quad \gotA\, [8 + 3\,\gotB^2 - 5\,\gotB^4 + 
\gotA^2\, (9\, \gotB^2 + 2\,\gotB^4-8)]\, \cos\gotf{}\, \sin\gotf{}\,\sin\beta,\nonumber\\     
&&c_{3,B,A,2,1}=-4\, \gotA\, \gotap\, \gotB\, \Lambda_1^4\, \cos\gotf{}\, \sin\beta,\\
 &&c_{3,B,A,2,2} = 
\Lambda_1^2\, [\gotA^2 (2\, \gotB^2-3) + 3 \Lambda_1^2] (\gotbp\, \cos\gotf{} - \cos\beta\, \sin(\gotf{}) + \\
&&\quad\quad\quad\quad \gotB\, [3\,\Lambda_1^2-\gotA^2\, (7+2\, \gotB^2 ) ]\, \cos\gotf{}\, 
\sin\gotf{}\, \sin\beta)),\nonumber \\
 && c_{3,B,A,3,1}  =-8\, \gotap\, \Lambda_1^6\, \cos\gotf{}\, \sin\beta,\\
 &&c_{3,B,A,3,2} =  -2\, \gotA\, \Lambda_1^4\, \{
 \cos\gotf{}\,[ \gotB\, \gotbp+(4 + 5\, \gotB^2)\,\sin\gotf{} \sin\beta]- \\
&&\quad\quad \quad\quad\quad  \gotB\, \cos\beta \sin\gotf{}\, \} ,    
\quad\quad\quad\quad\quad\quad  
c_{3,B,A,4,1}=0, \\
&&c_{3,B,A,4,2}=-6\, \Lambda_1^6 \,[ (\gotbp + \gotB\, \sin\gotf{}\, \sin\beta)\,\cos\gotf{}  
-\cos\beta\, \sin\gotf{}] , \\
&&\gotP_{3,B,A}(t)= \frac{\Delta_1}{24\, \Lambda_1^6}\,\sum_{i=1}^3\sum_{j=1}^2\,
c_{3,B,A,i,j}\,t^{i-1}\,r^{j-1};
\end{eqnarray}
\begin{eqnarray}
&& \Omega_{3,B,3}=\frac{ \Lambda_1\,  \Delta_1}
{\gotA\, \gotB + \Lambda_2 + \Lambda_1^2 t},\\
&& c_{3,B,B,3,1} = -4\, \gotA\, \gotap\, \gotB\, \Lambda_1^2\, \cos\gotf{}\, \sin\beta,\\ 
&&c_{3,B,B,3,2} =\gotB (5\, \Lambda_2^2-4 \Lambda_1^2)\, \cos\gotf{}\, \sin\gotf{} \sin\beta+\\
&&\quad\quad\quad\quad [4\, \Lambda_1^4 + 5\, \Lambda_2^2 - 4\, \Lambda_1^2 (1 + \Lambda_2^2)] \,(\gotbp\, \cos\gotf{} - \cos\beta\, \sin\gotf{})  ,\nonumber \\
&&\gotP_{3,B,B}(t)= \frac{\Lambda_2^2\,\arctan(\Omega_{3,B,3})}{4\, \Lambda_1^7}\,  
\sum_{j=1}^2\,c_{3,B,B,3,j}\,\, r^{j-1}.
\end{eqnarray}                                          
\subsubsection*{The  parallel case}
In this case, the CLD takes the form of equation \refeq{7.9} and this integral 
always is a linear combination of two integrals the integrands of which, as  it results from 
equations \refeq{7.1}, \refeq{7.8a}, \refeq{7.8b} and \refeq{7.5c},  have the  form 
\begeq\label{p.p}
\gotF_{p}(y,\varphi)\equiv\cA+\cB\,\cos\varphi+\cC\,\sin\varphi+\cD\,y
\endeq 
where constants $\cA,\,\cB,\,\cC,\,\cD$ depend on the contribution dictated by 
the inequality fulfillment. For instance, if we had to consider $\overline{\cL\p}(y)$, we 
would find: 
$$\cA=\gotA\p,\quad \cB=-\sqrt{r^2-h^2},\quad \cC=\gotBp\,\sqrt{r^2-h^2}\quad 
{\rm and} \quad  \cD=\gotBp.$$
The $y$-primitive of \refeq{p.p} is (setting the integration constant equal to zero)
\begeq
\overline{\gotF_{p}}(y,\varphi)\equiv(\cA+\cB\,\cos\varphi+\cC\,\sin\varphi)\,y+\cD\,y^2/2.
\endeq 
Each of the  end points of the $y$-integration domain has the general form: 
$\gotC+\gotD\,\sin\varphi $ with $\gotC$ and $\gotD$ suitable constants depending 
on the geometrical parameters. Hence, the explicit determination of integral \refeq{7.9} 
is certainly possible if we explicitly know the $\varphi$-primitive of 
$\overline{\gotF_{p}}(\gotC+\gotD\,\sin\varphi,\,\varphi)$. 
The explicit expression of this primitive reads
\begin{eqnarray}
&&{\gotP_{p}}(\varphi)= \frac{4\, \cA\, \gotC + 2\, \cD\, \gotC^2 + 2\, \cC\, \gotD + 
     \cD\, \gotD^2}{4}\, \varphi+
\quad\quad\quad\quad\quad\quad \\
&&\quad\quad\quad  \cB\,\gotC\, \sin\varphi -  
\frac{\gotD\, (2\, \cC + \cD\,\gotD)}{8} \sin(2 \varphi)-\nonumber \\
&&\quad\quad\quad  [\cC\, \,\gotC + (\cA +\cD\, \gotC)\, \gotD] \,\cos\varphi -
\frac{\cB\, \gotD}{4}\, \cos(2 \varphi).\nonumber  
\end{eqnarray} 
\vfill\eject
\subsection*{References}
\begin{description}
\vbox{
\item[\refup{1}] A. Guinier  and G. Fournet,  {\em Sall-Angle Scattering of 
X-rays}, New York: John Wiley,  (1955).
\item[\refup{2}] L.A. Feigin  and D.I.   Svergun, {\em Structure Analysis
by Small-Angle X-Ray and  Neutron Scattering}, New York: Plenum Press,  (1987).
\item[\refup{3}] S.W. Smith, {\em The Scientist and Engineer’s Guide to Digital 
Signal Processing} (2nd ed.). Los Angeles:California Technical Publishing, (1999). 
\item[\refup{4}] J. Serra,  {\em Image Analysis and Mathematical Morphology.} 
New York:Academic Press, (1982).
\item[\refup{5}] S.N. Chiu, D. Stoyan, W.S. Kendall and J.  Mecke, {\em 
Stochastic Geometryand its Applications}, 3rd ed. Chichester: Wiley,  (2013).
\item[\refup{6}] N. Wiener,  {\em  Acta Math.} {\bf 30}, 118-242, (1930).
\item[\refup{7}] S. Ciccariello, P. Riello and  A. Benedetti,, 
{\em  J. Appl. Cryst..} {\bf {51}}, 1404-1420, (2018). 
\item[\refup{8}] S. Ciccariello, G.   Cocco, A. Benedetti and  S. Enzo, 
{\em  Phys. Rev.} {\bf {B23}}, 6474-6485, (1981). 
\item[\refup{9}] G. Math\'eron, {\em Ensembles al\'eatoires et g\'éom\'étrie 
intégrale}. Les cahiers du Centre de Morphologie Mathématique 6. ENSMP, (1972). 
\item[\refup{10}] J. Goodisman, {\em  J. Appl. Cryst.} {\bf {13}}, 132-134,  (1980).
\item[\refup{11}] W. Gille,  {\em J. Appl. Cryst.} {\bf 32}, 1100-1104, (1999).
\item[\refup{12}] S. Ciccariello, {\em J. Appl. Cryst.} {\bf 38}, 97-106,  (2005).
\item[\refup{13}] S. Ciccariello, (2014). {\em J. Appl. Cryst.} {\bf {47}}, 1445-1448.
\item[\refup{14}] S. Ciccariello,  {\em  Acta. Cryst.} {\bf {A45}}, 86-99,  (1989).
\item[\refup{15}] W.  Gille, {\em Particle and Particle Systems
Characterization}, London: CRC,  (2014).
\item[\refup{16}] S. Ciccariello,  {\em J. Math. Phys.} {\bf 50}, 103527,  (2009).
\item[\refup{17}] R. Garcia-Pelayo,  {\em arXiv 1501.00292v1}, (2015).
\item[\refup{18}] R. Caccioppoli,  {\em Lezioni di Analisi Matematica.}  
Napoli: Treves. Vol. II, pagg. 175-180,  (1956). 
\item[\refup{19}] S. Ciccariello, {\em Fibres \& Text. East Eur.} {\bf 13}, 41-46,  (2005).}
\end{description}
\end{document}